\documentclass[%
 aip,jap,
 amsmath,amssymb,
 showpacs,
 reprint,%
superscriptaddress]{revtex4-1}

\usepackage{graphicx}
\usepackage{dcolumn}
\usepackage{bm}

\begin{document}

\title{A phonon scattering assisted injection and extraction based terahertz quantum cascade laser}

\author{E. Dupont}
\email{emmanuel.dupont@nrc-cnrc.gc.ca}
\affiliation{Institute for Microstructural Sciences, National Research Council, Ottawa, Ontario K1A0R6, Canada}
\author{S. Fathololoumi}
\affiliation{Institute for Microstructural Sciences, National Research Council, Ottawa, Ontario K1A0R6, Canada}
\affiliation{Department of Electrical and Computer Engineering, Waterloo Institute of Nanotechnology, University of Waterloo, 200 University Ave W., Waterloo, Ontario N2L3G1, Canada}%
\author{Z.R. Wasilewski}
\affiliation{Institute for Microstructural Sciences, National Research Council, Ottawa, Ontario K1A0R6, Canada}
\author{G. Aers}
\affiliation{Institute for Microstructural Sciences, National Research Council, Ottawa, Ontario K1A0R6, Canada}
\author{S. R. Laframboise}
\affiliation{Institute for Microstructural Sciences, National Research Council, Ottawa, Ontario K1A0R6, Canada}
\author{M. Lindskog}
\affiliation{Division of Mathematical Physics, Lund University, Box 118, 22100 Lund, Sweden}
\author{A. Wacker}
\affiliation{Division of Mathematical Physics, Lund University, Box 118, 22100 Lund, Sweden}
\author{D. Ban}
\affiliation{Department of Electrical and Computer Engineering, Waterloo Institute of Nanotechnology, University of Waterloo, 200 University Ave W., Waterloo, Ontario N2L3G1, Canada}
\author{H. C. Liu}
\affiliation{Key Laboratory of Artificial Structures and Quantum Control, Department of Physics, Shanghai Jiao Tong University, Shanghai 200240, China}
\date{\today}

\begin{abstract}
A novel lasing scheme for terahertz quantum cascade lasers, based on consecutive phonon-photon-phonon emissions per module, is proposed and experimentally demonstrated. The charge transport of the proposed structure is modeled using a rate equation formalism. An optimization code based on a genetic algorithm was developed to find a four-well design in the $\mathrm{GaAs/Al_{0.25}Ga_{0.75}As}$ material system that maximizes the product of population inversion and oscillator strength at 150 K. The fabricated devices using Au double-metal waveguides show lasing at 3.2 THz up to 138 K. The electrical characteristics display no sign of differential resistance drop at lasing threshold, which suggests - thanks to the rate equation model - a slow depopulation rate of the lower lasing state, a hypothesis confirmed by non-equilibrium Green's function calculations.
\end{abstract}

\maketitle

\section{\label{intro} Introduction}
Nearly a decade after the first demonstration of terahertz (THz) quantum cascade lasers (QCL)\cite{Kohler_Nat02}, the maximum operating temperature ($T_\mathrm{max}$) of these devices has reached 199 K, using a three-well resonant phonon design\cite{Fathololoumi_unpub}. The high temperature THz-QCLs ($T_\mathrm{max}\gtrsim~\mathrm{175~K}$) are mostly designed using resonant tunneling based injection and extraction of carriers from the lasing states in a $\mathrm{GaAs/Al_{0.15}Ga_{0.85}As}$ material system \cite{Belkin_OptExp_08,Kumar_186K_APL09,Belkin_SelTopQE09,Fathololoumi_unpub}. Despite the relatively slow progress on improving the $T_\mathrm{{max}}$, research efforts remain very intense to bring THz QCL into a temperature range achievable with thermo-electric coolers. This would open up a profusion of THz applications in many areas, including high speed communications, pharmacology, non-invasive cross sectional imaging, quality control, gas and pollution sensing, biochemical label-free sensing and security screening\cite{Tonouchi_Nat07}.

Several theoretical models have been employed to understand details of charge transport and optical gain in THz QCLs, including density matrix formalism \cite{Scalari_APL07,Kumar_DMmodel_PRB09,Dupont_PRB10}, non-equilibrium Green's function (NEGF) \cite{LeeWacker_PRB02,Kubis_PRB09_NEGFthzqcl,Schmielau_NEGF_APL09}, and Monte Carlo techniques \cite{Callebaut_APL03,Jirauschek_JAP09_MCthzqcl}. These models have identified several limitations of THz QCLs based on resonant tunneling injection (RT-QCL), which make it difficult for the lasers with photon energy $\hbar\omega$ to operate at temperatures $T>\hbar\omega/k_{B}$ ($k_{B}$ being Boltzmann constant). It is therefore necessary to find alternative designs with novel lasing schemes and/or investigate materials with lower electron effective mass \cite{Deutsch_APL10}. In this paper we focus on the first approach.

Yasuda\cite{Yasuda_APL09}, Kubis\cite{Kubis_APL10}, Kumar\cite{Kumar_NatPhy11} have also pointed out the limitations of RT-QCL and proposed a scheme formerly demonstrated in mid-infrared devices\cite{Yamanishi_OptExp_08}, for THz QCLs. In this scheme, carrier injection to the upper lasing state (ULS, also called level \emph{2}) is assisted by longitudinal optical (LO) phonon scattering, and is, hence, so-called scattering-assisted QCL (SA-QCL). This type of lasing scheme has recently proven its capability to significantly surpass the empirical $T_\mathrm{max}\sim~\hbar\omega/k_{B}$ limitation of RT-QCL: Kumar \emph{et al.} have demonstrated a four-well $\mathrm{GaAs/Al_{0.15}Ga_{0.85}As}$ SA-QCL at 1.8 THz with $T_\mathrm{max}=163~\mathrm{K}\sim~1.9\hbar\omega/k_{B}$\cite{Kumar_NatPhy11}. A similar design of SA-QCL consisting of five wells per module was also demonstrated at 4 THz and below 50 K in the lattice-matched InGaAs-AlInAs-InP material system\cite{Yamanishi_ITQW11}. In the following, we propose a novel four-well SA-QCL, where the lower lasing state (LLS, also called level \emph{1}) is depopulated without the mediation of resonant tunneling. The proposed design involves only four fairly confined states that are isolated from higher energy levels.

We recall the main drawbacks faced by the RT-QCL. The population inversion ($\Delta N/N_{\mathrm{tot}}$) of three-well resonant phonon based active regions is limited to 50\%. This is due to the existence of the injector level, \emph{i}, where carriers wait to get resonantly injected into the ULS. Ideally, in the coherent transport regime, the ULS holds as many carriers as the injector level, which is only half of the total available carriers. In a realistic case, with rather thick injection barrier and presence of scattering channels, the population inversion falls below 50\%. Moreover, as the injector state is populated, thermal backfilling to the lower lasing state (LLS) cannot be ignored at high temperatures\cite{NelanderWacker_APL08,Yasuda_APL09}. It was also predicted that the presence of injection and extraction tunneling couplings broadens the gain with a complicated electric field dependency\cite{Wacker_SPIE09_gainQCL,Scalari_DoubWell_OE10,Kumar_DMmodel_PRB09,Dupont_PRB10}, which makes it hard to predict the laser frequency. In order to inject carriers selectively to the ULS and not the LLS, the injection barrier should be chosen meticulously: thick enough to avoid the counterproductive (also called ``wrong'') injection of carriers to the LLS as well as to reduce the negative differential resistance (NDR) at the \emph{i}-LLS level alignment, and at the same time, thin enough to increase the laser's current density dynamic range. It is important to note that due to the small ULS and LLS energy spacing (small photon energy), the \emph{i}-ULS and \emph{i}-LLS level alignments happen at similar electric fields. The diagonality of the lasing transition was used as a design tool to decouple the intermediate tunneling resonances and the injection resonance, which occur before and at the design electric field respectiviely\cite{Kumar_186K_APL09,Fathololoumi_ITQW11}. The tradeoff in designing the injection tunneling barrier is even harder to satisfy for devices with smaller photon energy. As a consequence, the transport across the injection barrier may not become completely coherent, particularly when considering all existing scattering processes, and therefore the peak current becomes sensitive to variations of dephasing scattering during tunneling\cite{Kumar_NatPhy11}.

\begin{figure}[t!]
\centering
\includegraphics[width=3in]{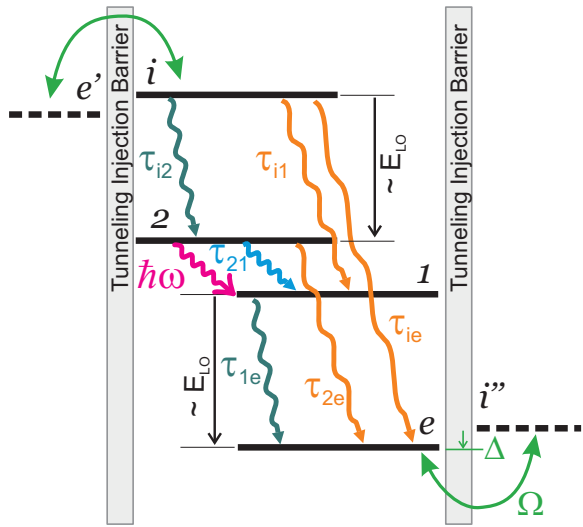}
\caption{(Color online) Schematic diagram of a scattering assisted QCL active region based on a ``phonon-photon-phonon'' configuration.}
\label{fig_SchematicSAQCL}
\end{figure}

Refs.~\onlinecite{Yasuda_APL09}, \onlinecite{Kubis_APL10} and \onlinecite{Kumar_NatPhy11} provide comprehensive discussions of how SA-QCL designs address the shortcomings of the RT-QCL. Figure \ref{fig_SchematicSAQCL} illustrates schematically the operation principle of the SA-QCLs, with consecutive ``phonon-photon-phonon'' emissions per module. In energy ascending order, the states are named \emph{e} (for extraction from LLS), \emph{1}, \emph{2} and \emph{i} (for injection to ULS). Such design schemes rely on an injection energy state, \emph{i}, lying about one LO-phonon energy above the ULS. The carriers are injected from the upstream extraction level, \emph{e'}, to the next injection level \emph{i} by resonant tunneling. Since the electric field at \emph{e'}-\emph{i} alignment is much larger than that for \emph{e'}-\emph{2} and \emph{e'}-\emph{1} alignments, the injection resonance \emph{e'}-\emph{i} is well decoupled from the last two resonances and a large coupling strength, $\Omega_{ei}$, can be used in the design. Therefore, it is possible to have thin tunneling barriers, while preserving the pumping selectivity to the ULS, provided the diagonality between the lasing states is adjusted to minimize the relaxation from \emph{i} to \emph{1}. As a result, this type of design relies on a large diagonality between lasing states, hence the occurrence of photon emission assisted tunneling through a rather thick ``radiative barrier''\cite{Kumar_NatPhy11}. By choosing a thin injection barrier, one can ensure coherent transport and subsequent accumulation of carriers on the state with longest lifetime - preferably the ULS. This would also leave a small concentration of carriers on the short lifetime injection state \emph{i} and extractor state \emph{e'}, which reduces the effect of backfilling to state \emph{1'}. The coherent transport through the barrier, additionally, makes current and population inversion relatively insensitive to variations of dephasing processes. As shown in Figure~\ref{fig_SchematicSAQCL}, for depopulating the LLS we simply propose to rely on LO-phonon emission scattering, an extraction scheme similar to that of Refs.~\onlinecite{Kumar_2well_APL09,Scalari_DoubWell_OE10} in two-well RT-QCLs. Our design is similar to the one proposed in Ref.~\onlinecite{Kubis_APL10}. Alternatively, one could use the mediation of tunneling to an excited state in resonance with LLS before LO-phonon emission scattering to the extraction state, a configuration that could be named ``phonon-photon-tunnel-phonon''\cite{Kumar_NatPhy11,Yamanishi_ITQW11}. One could also skip the depopulation by LO-phonon emission and make the LLS acting as a state that feeds the injection state of the next module by resonant tunneling, an extraction controlled ``phonon-photon'' configuration which offers the advantage of a lower voltage defect\cite{Wacker_APL10}.

In this work, we focus on ``phonon-photon-phonon'' scheme and evaluate the lasing states (and gain profile) that are not perturbed by injection and/or extraction tunnelings. In such a scheme, the relative diagonality between the four states should be carefully adjusted to insure fast injection to ULS and extraction from LLS, while minimizing the wrong injection \emph{i}$\rightarrow$\emph{1} and extraction \emph{2}$\rightarrow$\emph{e} channels and at the same time maintaining a sufficient oscillator strength between the lasing states. In the next section, we will discuss the key figures of merit, derived from rate equations and used for design optimization.

\section{\label{rateequations} Rate Equations}
The key figures of merit, such as population inversion and transit time of a ``phonon-photon-phonon'' design, can be derived analytically from rate equations, thanks to the operational simplicity of such SA-QCL designs. We denote the scattering rate from state $l$ to $m$ as  $g_{lm}$ and the inverse of $g_{lm}$ as $\tau_{lm}$. The tunneling time across the injection barrier can be written as $\tau_{\mathrm{tun}}=(1+\Delta^2\tau_{\|}^2)/2\Omega^2\tau_{\|}$, where $\hbar\Delta$ and $\hbar\Omega$ are, respectively, the detuning energy and tunneling coupling strength between the states \emph{e} and \emph{i}, and $\tau_{\|}$ denotes the phase coherence time constant\cite{Khurgin_rateequation10}. For the rate equation model, the basis of states is taken from the eigenstates of ``isolated modules'', as if the injection barrier were very thick. Simplified analytical forms can be obtained when backscattering is neglected (i.e. low temperature): $g_{ml}\ll~g_{lm}$, where $E_l>E_m$. The total down scattering rates from levels \emph{i} and \emph{2} are defined as $g_{i}=g_{i2}+g_{i1}+g_{ie}$ and
$g_{2}=g_{21}+g_{2e}$, respectively. For the sake of simplicity, we define $g_1=g_{1e}$. In the absence of stimulated emission, i.e. below threshold ($J<J_{\mathrm{th}}$), the population inversion normalized to the total number of carriers per period reads\\
\begin{eqnarray}
\Delta\rho&=&\frac{\widetilde{\tau}_{2\,\mathrm{eff}}}{\tau_{\mathrm{tun}}+\tau_{\mathrm{tr}}^{<}}\label{Eq_Dn},
\end{eqnarray}\\
where $\widetilde{\tau}_{2\,\mathrm{eff}}$ is called \emph{the modified effective lifetime} of ULS and is defined as\\
\begin{eqnarray}
\widetilde{\tau}_{2\,\mathrm{eff}}=\frac{g_{i2}(g_{1}-g_{21})-g_{2}g_{i1}}{g_{1}g_{2}g_{i}},\label{Eq_modtau2eff}
\end{eqnarray}
and $\tau_{\mathrm{tr}}^{<}$ is the transit time below threshold defined as \\
\begin{eqnarray}
\tau_{\mathrm{tr}}^{<}&=&\frac{g_{i2}(g_{1}+g_{21})+g_{2}(2g_{1}+g_{i1})}{g_{1}g_{2}g_{i}}\label{Eq_transWOL}.
\end{eqnarray}\\
The normalized populations on states \emph{i} and \emph{e} can be written simply as\\
\begin{eqnarray}
\rho_{ii}&=&\frac{\tau_{i}}{\tau_{\mathrm{tun}}+\tau_{\mathrm{tr}}^{<}}\label{Eq_nu}
\end{eqnarray}\\
and
\begin{eqnarray}
\rho_{ee}&=&\frac{\tau_{\mathrm{tun}}+\tau_{i}}{\tau_{\mathrm{tun}}+\tau_{\mathrm{tr}}^{<}},\label{Eq_ni}
\end{eqnarray}\\
respectively. The current density below threshold was estimated in the middle of the tunneling barrier. Assuming this layer to be crossed only by tunneling currents, the current density can be written simply as\\
\begin{eqnarray}
J^{<}=qN_s\frac{\rho_{ee}-\rho_{ii}}{\tau_{\mathrm{tun}}}=q\frac{N_s}{\tau_{\mathrm{tun}}+\tau_{\mathrm{tr}}^{<}},\label{Eq_JWOL}
\end{eqnarray}\\
where $N_s$ is the two-dimensional carrier density per module. In the ideal case, where leakages $g_{i1}$, $g_{ie}$ and $g_{2e}$ are negligible, the transit time and population inversion converge towards intuitive expressions: $\tau_{\mathrm{tr}}^{<}\approx~2\tau_{i2}+\tau_{21}+\tau_{1}$ and $\Delta\rho\approx~(\tau_{21}-\tau_{1})/(\tau_{\mathrm{tun}}+\tau_{\mathrm{tr}}^{<})$, respectively. If the intersubband relaxation time between the lasing states, $\tau_{21}$, is much longer than the injection and extraction lifetimes, and if the tunneling time is much shorter than the transit time, the population inversion could approach 100\%. Eq.~\ref{Eq_ni} also suggests that to avoid accumulation of carriers behind the tunneling barrier, the transport should be coherent ($\tau_{\mathrm{tun}}\ll~\tau_{i}$ or equivalently, $2\Omega^{2}\tau_{\|}\tau_{i}\gg1$).

In the presence of stimulated emission ($J>J_{\mathrm{th}}$), the population inversion is clamped at $\Delta\rho_{\mathrm{th}}$, which is fixed by cavity loss and the gain cross-section. One can show that the current density above threshold is\\
\begin{eqnarray}
J^{>}=qN_{s}\frac{1+\Delta R_{\mathrm{th}}/R_{\mathrm{th}}}{\tau_{\mathrm{tun}}+\tau_{\mathrm{tr}}^{>}},\label{Eq_JWIL}
\end{eqnarray}\\
where $\tau_{\mathrm{tr}}^{>}$ is the transit time above threshold and is approximately $4\tau_{i}$, assuming efficient and similar injection and depopulation rates ($g_{i2}\approx g_1$). More precisely, the transit time above threshold becomes\\
\begin{eqnarray}
\tau_{\mathrm{tr}}^{>}=2\tau_{i}\left(1+\frac{g_{i2}+ g_{i1}}{g_{1}+g_{2e}}\right).\label{Eq_transWIL}
\end{eqnarray}\\
The fractional change in differential resistance at threshold, $\Delta R_{\mathrm{th}}/R_{\mathrm{th}}$, is also calculated as\\
\begin{eqnarray}
\frac{\Delta R_{\mathrm{th}}}{R_{\mathrm{th}}}=-\frac{R_{\mathrm{th}}^{>}-R_{\mathrm{th}}^{<}}{R_{\mathrm{th}}^{<}}=\Delta \rho_{\mathrm{th}}\frac{g_{1}-g_{2e}}{g_{1}+g_{2e}}.\label{Eq_DRoR}
\end{eqnarray}\\
A discontinuity of differential resistance is expected at threshold, as the current below threshold depends on the intersubband relaxation between lasing states, $\tau_{21}$ (see Eq.~\ref{Eq_modtau2eff}), and above threshold this relaxation is governed by stimulated emission. From Eqs.~\ref{Eq_Dn}, \ref{Eq_JWOL} and \ref{Eq_JWIL}, the current density dynamic range gives a condensed form of \\
\begin{eqnarray}
\frac{J_{\mathrm{max}}}{J_{\mathrm{th}}}&=&\frac{1+\Delta R_{\mathrm{th}}/R_{\mathrm{th}}}{\Delta\rho_{\mathrm{th}}}
\frac{2\Omega^2\tau_{\|}\widetilde{\tau}_{2\,\mathrm{eff}}}{1+2\Omega^2\tau_{\|}\tau_{\mathrm{tr}}^{>}},\nonumber\\
\lim_{\mathrm{coh. trans.}}\left(\frac{J_{\mathrm{max}}}{J_{\mathrm{th}}}\right)&=&\frac{1+\Delta R_{\mathrm{th}}/R_{\mathrm{th}}}{\Delta\rho_{\mathrm{th}}}\frac{\widetilde{\tau}_{2\,\mathrm{eff}}}{\tau_{\mathrm{tr}}^{>}}.\label{Eq_DymRange}
\end{eqnarray}\\
The second expression shows the importance of using coherent transport across the tunneling barrier in order to maximize the dynamic range, a condition reached when $2\Omega^2\tau_{\|}\tau_{\mathrm{tr}}^{>}\gg 1$, or approximatively when $8\Omega^2\tau_{\|}\tau_{i}\gg 1$. The differential internal efficiency of the laser is obtained as\\
\begin{eqnarray}
\eta=\frac{g_{i2}(g_{1}-g_{21})-g_{2}g_{i1}}{g_{i}(g_{1}+g_{2e})}.\label{Eq_eta1}
\end{eqnarray}\\
If we define the \emph{effective lifetime} of level \emph{2} by the standard expression, $\tau_{2\,\mathrm{eff}}=\tau_2(1-\tau_{1}/\tau_{21})$, more intuitive expressions for the modified effective lifetime and the internal efficiency can be obtained as,\\
\begin{eqnarray}
\widetilde{\tau}_{2\,\mathrm{eff}}&=&\frac{g_{i2}}{g_{i}}\tau_{2\,\mathrm{eff}}-\frac{g_{i1}}{g_{i}}\tau_{1}\;\mathrm{and}\label{Eq_modtau2eff_bis}\\
\eta&=&\frac{\widetilde{\tau}_{2\,\mathrm{eff}}}{\tau_{2\,\mathrm{eff}}+\tau_{1}},\label{Eq_eta2}
\end{eqnarray}\\
respectively. Eq.~\ref{Eq_modtau2eff_bis} shows how the effective ULS lifetime is modified by the leakages channels \emph{i}$\rightarrow$\emph{1} and \emph{i}$\rightarrow$\emph{e} that are shunting the ``correct'' trajectory of carriers inside a module from \emph{i}$\rightarrow$\emph{2}$\rightarrow$\emph{1}$\rightarrow$\emph{e}. The fractional change in differential resistance at threshold is closely related to the internal efficiency; it can be re-written versus $\tau_{2\,\mathrm{eff}}$ as\\
\begin{eqnarray}
\frac{\Delta R_{\mathrm{th}}}{R_{\mathrm{th}}}&=&\rho_{\mathrm{th}}\frac{\tau_{2\,\mathrm{eff}}+
\tau_{1}\frac{g_{21}-g_{2e}}{g_{2}}}{\tau_{2\,\mathrm{eff}}+\tau_{1}}, \label{Eq_DRoRbis}
\end{eqnarray}\\
an expression which shows similarities with Eq.~\ref{Eq_eta2}. From Eq.~\ref{Eq_DRoR}, it appears that the discontinuity in differential resistance would vanish if the depopulation rate, $g_1$, is inefficient and close to the wrong depopulation channel, $g_1\approx g_{2e}$. Nevertheless, in such adverse conditions, population inversion is still possible, if the transition rate between lasing state, $g_{21}$, is lower than $g_{21\,\mathrm{lim}}=g_{2e}\frac{g_{i2}-g_{ie}}{g_{i2}+g_{ie}}$. In this situation, the modified effective lifetime, $\widetilde{\tau}_{2\,\mathrm{eff}}$, would be small $\sim (g_{21\,\mathrm{lim}}-g_{21})/2g_{2e}^2$ (for $g_{21}\lesssim g_{21\,\mathrm{lim}}$), implying the efficiency of the laser will be impeded. Of course, the transit time above threshold would be much longer than $4\tau_{i}$. In the next section, we describe how the rate equation model was employed to design the first iteration of SA-QCL based on the ``phonon-photon-phonon'' scheme.

\section{\label{Design} Design of structure}
For the first iteration of the ``phonon-photon-phonon'' design, we optimized a figure of merit defined as the product of population inversion, oscillator strength and inverse of the superperiod length. We chose to maximize this figure of merit using four quantum wells in the $\mathrm{GaAs/Al_{0.25}Ga_{0.75}As}$ material system at 21 kV/cm and a lattice temperature of 150 K. We assumed a Maxwell-Boltzmann distribution for the carriers in all subbands, with a characteristic temperature of $T_e=200~\mathrm{K}$. For the first iteration, we assumed that transport is dominated by LO-phonon scattering, and neglected electron-impurity and interface roughness (IR) scattering, which can be a serious shortcoming. The calculation included both forward and backward scattering channels. A more accurate figure of merit would include several scattering potentials acting on th population rate and dephasing of the lasing transition. To avoid leakage channels to higher energy states, we deliberately opted for a rather thick injection barrier of 44~$\mathrm{\AA}$. Within this model, a genetic algorithm based optimization approach was used to find the optimum thicknesses of four quantum wells and three barriers (the injection tunneling barrier thickness being fixed). The energy spacing between the four levels were unbound during the optimization process.

\begin{figure}[t!]
\includegraphics[width=5in]{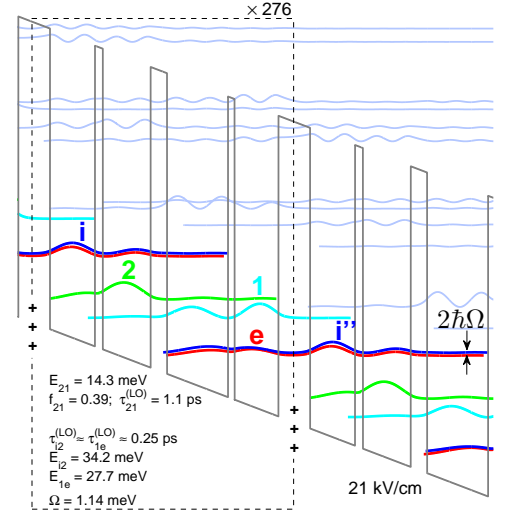}
\caption{(Color online) Conduction band diagram and the moduli squared of wavefunctions at 21 kV/cm of the designed THz QCL active region based on ``phonon-photon-phonon'' scheme. The ``+'' signs denote the ionized impurities used for doping. The LO-phonon emission assisted scattering times $\tau_{lm}^{(\mathrm{LO})}$ are given for the initial kinetic energy $E_{\mathrm{LO}}-E_{lm}$.}
\label{fig_Band_diagram}
\end{figure}

The optimization resulted in the solution schematically shown in Figure \ref{fig_Band_diagram}, at 21 kV/cm. Starting from the injection barrier, it consists of four wells and barriers with the layer thicknesses of \textbf{44}/62.5/\textbf{10.9}/66.5/\textbf{22.8}/84.8/\textbf{9.1}/61 $\mathrm{\AA}$ - the barriers are indicated in bold fonts. The injection barrier is delta-doped with Si to $3.25\times10^{10}~\mathrm{cm}^{-2}$, at the center. Interestingly, with the discussed figure of merit, which only includes LO-phonon scattering and the assumption of the same electronic temperature of 200 K in all subbands, the genetic algorithm converged towards a low energy spacing between the LLS and the extraction level, $E_{1}-E_{e}=27.7~\mathrm{meV}$, i.e. 9 meV lower than th bulk GaAs LO-phonon energy. This suggests that, within our simplified model, the depopulation rate is thermally activated (with a 9 meV activation energy), which impedes population inversion at low temperature. The electrostatic energy per module is 76 meV that is approximately twice the phonon energy, as proposed by Kubis \textit{et al.}, to favor carrier thermalization within one module\cite{Kubis_APL10}. With the choice of thick injection barrier, the tunneling coupling strength is rather weak ($\hbar\Omega$=1.14 meV), comparing with 1.5 meV from the first demonstrated THz SA-QCL in Ref.~\onlinecite{Kumar_NatPhy11}.

\begin{table}[t!]
  \caption{\label{tabletaus}Scattering lifetimes (including LO-phonon, IR and e-impurity) in ps at 20 K and 150 K lattice temperatures. The IR scattering times are calculated with a correlation length of $\Lambda_{IR} =6.5~\mathrm{nm}$ and a mean roughness height of $\Delta_{IR}=2.83$~$\mathrm{\AA}$. In the fourth column, e-impurity scattering times are reported. In the last column the total backscattering lifetimes at 150 K are reported. Values are not reported for very long $\tau_{lm}>400~\mathrm{ps}$. In all subbands, electrons were distributed according to Maxwell-Boltzmann statistics with a temperature, $T_e$, 50 K higher than lattice.}
  \begin{ruledtabular}
    \begin{tabular}{c|ccc|ccc}
    &\multicolumn{3}{c|}{20 K}&\multicolumn{3}{c}{150 K}\\

    $l-m$&$\tau_{lm}$&$\tau_{lm}^{(\mathrm{IR})}$&$\tau_{lm}^{(\mathrm{imp})}$&$\tau_{lm}$&$\tau_{lm}^{(\mathrm{IR})}$&$\tau_{ml}$\\
    \hline
\emph{1}-\emph{e}&0.7&1.7&11.8&0.41&2&3\\
\emph{2}-\emph{e}&2.6&18.4&-&2.8&21.2&53\\
\emph{2}-\emph{1}&2.4&3.2&12.7&1.7&3.9&4.5\\
\emph{i}-\emph{e}&9&20.6&-&9.2&22.1&-\\
\emph{i}-\emph{1}&2.4&7.8&-&2.7&8.8&65\\
\emph{i}-\emph{2}&0.36&1.25&26.3&0.33&1.46&3.7\\

  \end{tabular}
   \end{ruledtabular}
\end{table}

In the conduction band diagram shown in Figure~\ref{fig_Band_diagram} at 21 kV/cm, the four main wavefunctions are completely isolated from higher energy states. For simplicity, when the wavefunctions are solved for several modules (opposite case to ``isolated modules'') we keep the already-discussed notation of states \emph{e}, \emph{1}, \emph{2} and \emph{i}, even when the wavefunctions are hybridized from a pair of states that are in resonance, like levels \emph{e} and \emph{i''} in Figure~\ref{fig_Band_diagram}. The optimization converged towards excellent wavefunction overlaps between \emph{i}-\emph{2} and \emph{1}-\emph{e}, which led to short LO-phonon emission scattering times of 0.25 ps for both transitions at initial kinetic energy $E_{\mathrm{LO}}-E_{i2,1e}$. The thick ``radiative barrier'' (22.8 $\mathrm{\AA}$) and the strong diagonality between the lasing states lead to an oscillator strength of 0.39 at 3.46 THz and a LO-phonon emission scattering time $\tau_{21}^{(\mathrm{LO})}\sim 1.1~\mathrm{ps}$ at 22.4 meV initial kinetic energy ($E_{\mathrm{LO}}-E_{21}$).  Since the four wavefunctions are fairly diagonal to each other and isolated from higher subbands, the free carrier absorption of the active region is probably small\cite{Carosella_FCA_arXiv}. This type of design would need to be experimentally optimized versus the doping density. Without the concern of free carrier absorption, the doping could be increased; for instance electron-electron scattering could favor carrier thermalization and potentially improve device performance but on the other hand, e-impurity scattering could increase the gain linewidth, particularly at high temperatures\cite{NelanderWacker_APL08}. Even though the optimization process was performed with only LO-phonon scattering, the subsequent simulations include LO-phonon, IR and e-impurity scattering potentials. For the sake of comparison, the scattering lifetimes of the six channels, shown in Figure \ref{fig_SchematicSAQCL}, are listed in Table~\ref{tabletaus}. As expected, with the assumption of Maxwell-Boltzmann carrier distributions, the total scattering time for depopulation, $\tau_{1e}$, is shorter at elevated temperatures, which is due to the 9 meV activation energy of LO-phonon scattering. The IR scattering is expected to give a large contribution between vertical states (i.e. \emph{i}-\emph{2} and \emph{1}-\emph{e}), but we find it is also not negligible between the two diagonal lasing states\cite{Kubis_APL10}. The IR scattering contributes 76\% (44\%) to the total scattering rate $g_{21}$ at 20 K (150 K). In the next optimization iteration, it would be important to include IR scattering in the figure of merit. Several wafers based on this simple design with different growth conditions could be tested in order to elucidate the role of IR scattering in THz QCL, a subject which is debated in the litterature\cite{Kubis_PRB09_NEGFthzqcl,Schmielau_NEGF_APL09,Scalari_THZQCL_Hfield_PRB07}. A more elaborate figure of merit could also include the gain bandwidth, estimated by various scattering potentials (IR scattering is likely to contribute the most\cite{Unuma_IR_APL01,Tsujino_APL05}). These refined models would adjust the relative diagonality between the four states. Table~\ref{tabletaus} shows that electron-impurity scattering gives a small contribution.

\begin{figure}[t!]
\includegraphics[width=3.2in]{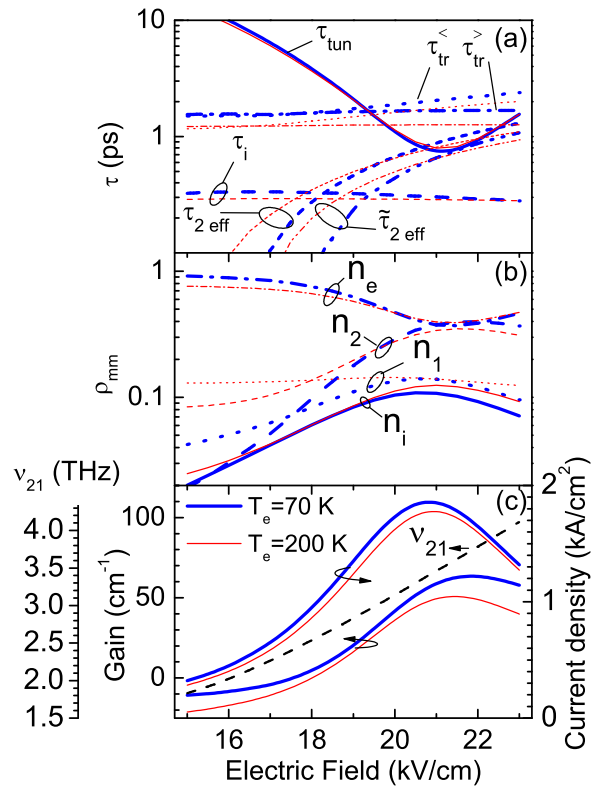}
\caption{(Color online) Results of simulations based on a rate equation model performed on the structure proposed in Figure~\ref{fig_Band_diagram}. a) Different characteristic times at 20 K (thick lines) and 150 K (thin lines)- 70 K and 200 K for electrons, respectively. As discussed in section~\ref{rateequations}, $\tau_{\mathrm{tun}}$ is tunneling time (solid line); $\tau_{\mathrm{tr}}^{<}$ (dot line) and $\tau_{\mathrm{tr}}^{>}$ (dash dot line) are transit times before and after threshold; $\tau_{i}$ is injection state lifetime (dash line); $\tau_{2\,\mathrm{eff}}$ is standard effective lifetime of ULS (short dash line); and $\widetilde{\tau}_{2\,\mathrm{eff}}$ is modified effective lifetime (dash dot line). b) Normalized populations of the four states at 20 and 150 K lattice temperatures. c) Current density, lasing frequency and optical gain versus electric field at 20 and 150 K lattice temperatures.}
\label{fig_SimuVsField}
\end{figure}

The current density, optical gain, populations and different characteristic lifetimes were calculated within the rate equation model. The results versus electric field at two temperatures (20 K and 150 K) are shown in Figure~\ref{fig_SimuVsField}. In the simulation, the coherence time constant, $\tau_{\|}$, has a pure dephasing time component, $\tau^\ast=0.35$~ps, which is assumed to be temperature independent for simplicity. The optical gain was estimated by assuming a constant intersubband linewidth of 1 THz. The different lifetimes, as introduced in section~\ref{rateequations}, are plotted in Figure~\ref{fig_SimuVsField}-a. We can see the tunneling time, $\tau_{\mathrm{tun}}$ is slightly shorter than the transit time above threshold, $\tau_{\mathrm{tr}}^{>}$ (by a factor $\approx 2.2$ at 20 K), meaning that the current dynamic range is not optimized (see Eq.~\ref{Eq_DymRange}). The tunneling time is substantially longer than the injection state lifetime, $\tau_{i}$, by a factor of 2.5, which indicates transport is not coherent through the barrier. As a result, at the design electric field, the populations on the states \emph{e} and \emph{i} are very different. In other words, carriers tend to accumulate on the extraction state, and hence promote backfilling to the LLS (see Eq.~\ref{Eq_ni}). Figure~\ref{fig_SimuVsField}-b illustrates the effect of backfilling on populations $\rho_{11}$, $\rho_{ee}$ and $\rho_{22}$, particularly at electric fields below 21 kV/cm. At the design electric field, the effect of backfilling at high temperature is mitigated by the faster relaxation between laser states $\tau_{21}$ and depopulation $\tau_{1e}$. At the design electric field, the population of level \emph{i} increases with temperature, as the transit time $\tau_{\mathrm{tr}}^{<}$ for carriers contributing to current decreases, while its lifetime remains unchanged. As shown in Figure~\ref{fig_SimuVsField}-c, the slight decrease in current density at higher temperature is related to backfilling to the LLS, meaning that less carriers participate in the transport.

\begin{figure}[t!]
\includegraphics[width=2.9in]{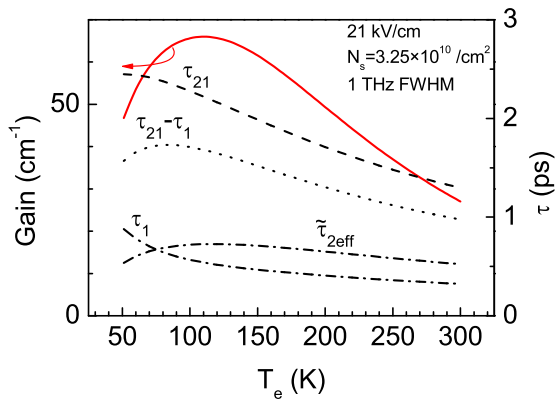}
\caption{(Color online) Left axis: peak gain at 21 kV/cm versus the electronic temperature from simulations based on rate equation model. Right axis: intersubband recombination time between lasing states, $\tau_{21}$, and depopulation time constant, $\tau_{1e}$, the difference $\tau_{21}-\tau_{1e}$, and the modified effective lifetime of ULS, $\widetilde{\tau}_{2\,\mathrm{eff}}$.}
\label{fig_SimuVsT}
\end{figure}

The peak gain value at 21 kV/cm was simulated versus temperature and the results are displayed in Figure~\ref{fig_SimuVsT}. According to the model presented in section~\ref{rateequations}, a gain peak value of more than $40~\mathrm{cm}^{-1}$ is maintained up to $T_{e}=230~\mathrm{K}$. The surprising rise of gain from low temperature up to $T_{e}=110~\mathrm{K}$ originates from the assumption of Maxwell-Boltzmann distributions. This is due to the temperature-induced enhancement of the depopulation rate (\emph{1}$\rightarrow$\emph{e}) being greater than that of the intersubband relaxation between the lasing states (\emph{2}$\rightarrow$\emph{1}), at low temperatures. Indeed, the latter has an activation energy of 22.4 meV by LO-phonon scattering (versus 9 meV for $g_{1e}$) and involves a pair of states that are more diagonal than the pair \emph{1}-\emph{e}. Consequently at low temperature, $g_{21}$ has a stronger contribution from IR scattering than $g_{1e}$, which is harder to change by LO-phonon scattering (see Table~\ref{tabletaus} and right axis of Figure~\ref{fig_SimuVsT}). We can, further, see from the Figure~\ref{fig_SimuVsT} that the temperature dependance of $\tau_{21}-\tau_{1e}$ and the modified effective lifetime, $\widetilde{\tau}_{2\,\mathrm{eff}}$, are mirrored in the gain. In the next section we will describe the experimental results obtained with the structure proposed in Figure~\ref{fig_Band_diagram}.

\section{\label{ExpResults} Experimental results}
The design presented in the Figure~\ref{fig_Band_diagram} was grown on a semi-insulating GaAs substrate by molecular beam epitaxy with 276 repeats to obtain a 10 $\mu\mathrm{m}$ thick active region. The active region was sandwiched between a bottom stack 100 nm of $\mathrm{3\times10^{18}\;cm^{-3}}$ $\mathrm{n^{+}}$ GaAs followed by 20 nm of non-intentionally doped GaAs spacer and a top stack of $\mathrm{8\times10^{17}/5\times10^{18}\;cm^{-3}}$ (40 nm/50 nm) $\mathrm{n^{+}}$ GaAs layers followed by 10 nm of $\mathrm{5\times10^{19}\;cm^{-3}}$ $\mathrm{n^{+}}$ GaAs layer and capped with 3 nm of low-temperature grown GaAs. The lower doping concentration of the bottom contact layer ($\mathrm{3\times10^{18}\;cm^{-3}}$) and the 20 nm spacer were meant to limit diffusion of Si dopants to the first period, which could induce intermixing between the first layers of the active region. Indeed, by transmission electron microscope imaging, we observed blurred interfaces on the first three barriers of the structures that were grown with 100 nm of $\mathrm{5\times10^{18}\;cm^{-3}}$ bottom $\mathrm{n^{+}}$ GaAs followed by 10 nm of GaAs spacer\cite{Wasilewski_1stperiodQCL_unpub}. Special emphasis was put on minimizing the drift of fluxes for Ga and Al during this long growth process. The X-ray diffraction rocking curve could be fitted perfectly with nominal parameters, with no extra broadening of satellites peaks, confirming the excellent stability of the growth rates (better than $0.5\%$), throughout the active region. The wafers were processed into THz QCL structures with Au double metal waveguides. Devices have the following dimensions: $\mathrm{\sim144\;\mu m}$ wide ridges with $\mathrm{\sim120\;\mu m}$ wide top Ti/Au metallization forming a Schottky contact, and $\mathrm{\sim1\;mm}$ long Fabry-Perot resonator. An In-Au wafer bonding technique was used \cite{Williams_MMwaveguide_APL03,Fathololoumi_SST11} and the ridges were fabricated by reactive-ion etching. The laser bars were indium soldered (epi-layer side up) on silicon carriers and then mounted in a He closed-cycle cryostat for measurements.

\begin{figure}[t!]
\includegraphics[width=3.3in]{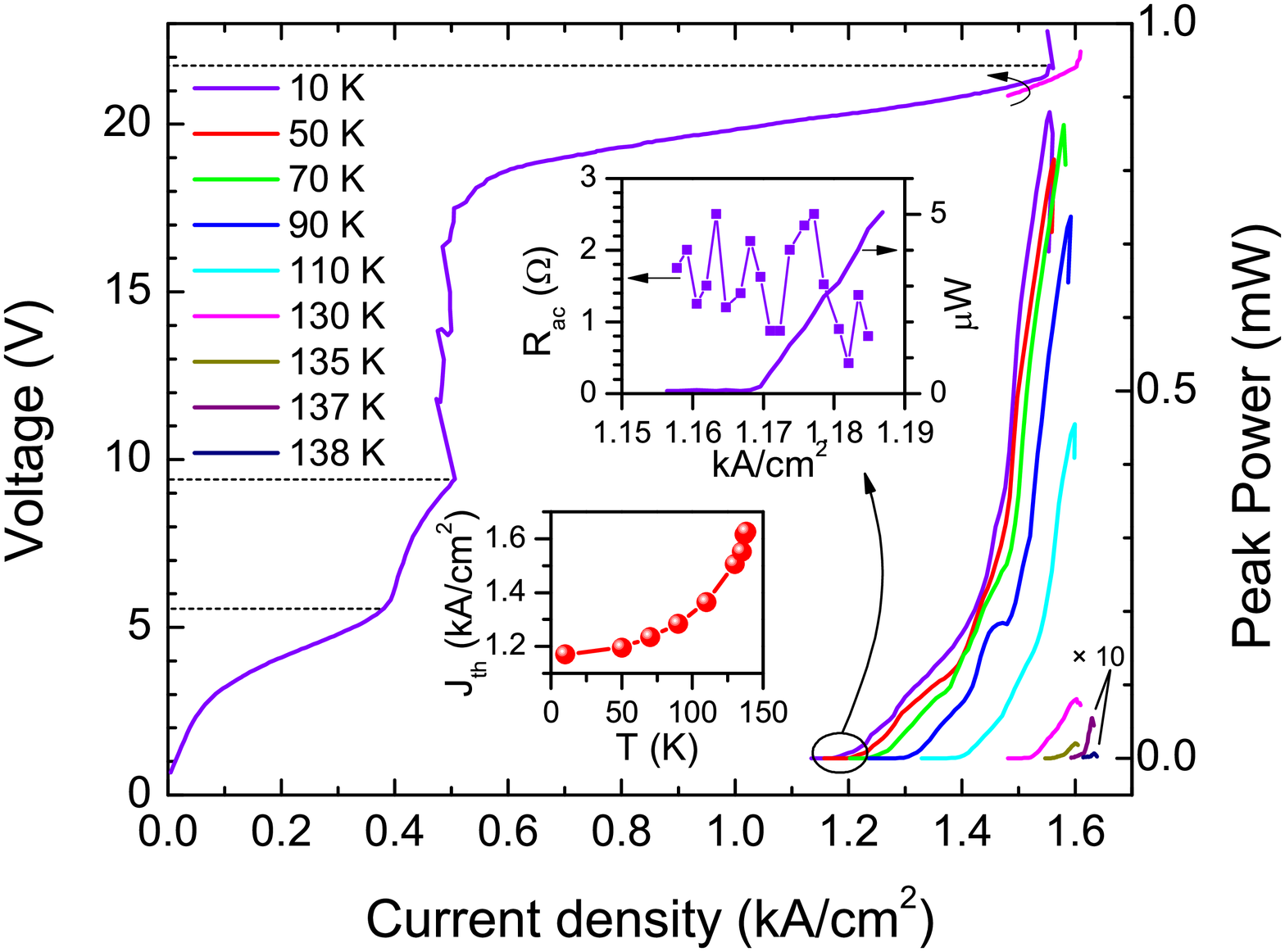}
\caption{ (Color online) Right axis: collected THz light (optical output power) versus current density curves for Au double-metal THz QCLs based on the ``phonon-photon-phonon'' lasing scheme at different heat sink temperatures. Left axis: voltage (V) versus current density (J) at 10 K and 130 K. The dashed horizontal lines highlight the position of resonances in the V-J plot. The devices are $\mathrm{\sim144\;\mu m}$ wide, 1 mm long and are fabricated using TiAu metal contacts. The bias is applied in pulsed mode (pulse width = 250 ns, repetition rate = 1 kHz). The top inset shows a zoom of THz light and differential resistance versus current density around threshold at 10 K. The bottom inset shows the threshold density versus temperature.}
\label{fig_LIV}
\end{figure}

Figure~\ref{fig_LIV} shows the light(L)-current density (J)-voltage (V) characteristics of the fabricated device from 10 K to 138 K in pulsed mode, with pulse duration of 250 ns and repetition rate of 1 kHz. The device showed a $T_\mathrm{max}$ of 138 K at a current density of $\mathrm{1.65\;kA/cm^{2}}$. At 10 K, the threshold current density is $\mathrm{1.17\;kA/cm^{2}}$ and the maximum current density is $\mathrm{1.55\;kA/cm^{2}}$ resulting in a small ratio of $J_{\mathrm{max}}/J_{\mathrm{th}}=1.325$. The laser action starts at 20.25 V and stops at 21.8 V where a NDR is observed. By subtracting the 0.8 V Schottky voltage drop on the top contact\cite{Fathololoumi_SST11}, the NDR voltage corresponds to the design electric field of 21 kV/cm. A zoom-view of the differential resistance, $\mathrm{R_{ac}}$, versus current density around threshold at 10 K (inset of Figure~\ref{fig_LIV}) did not show a clear discontinuity at threshold. Several measurements around threshold with different voltage steps were attempted but no clear change in the slope of the J-V characteristics could be detected. Analyzing this observation with the model discussed in section~\ref{rateequations} suggests that the depopulation rate \emph{1}$\rightarrow$\emph{e} is not as efficient as expected. Moreover, according to Eq.~\ref{Eq_DymRange}, the ratio $J_{\mathrm{max}}/J_{\mathrm{th}}$ can be approximated by $\widetilde{\tau}_{2\,\mathrm{eff}}/\Delta\rho_{\mathrm{th}}\tau_{\mathrm{tr}}^{>}$. The vanishing discontinuity of differential resistance is consistent with the low output power of the laser. It suggests that the effective lifetime of the ULS is much shorter than expected, and consequently reduces the internal efficiency of the laser (see Eq.~\ref{Eq_eta2}). The slow rate of depopulation could be related to the $E_{1e}$ energy spacing, being 9 meV smaller than the GaAs LO-phonon energy. As the LO-phonon emission scattering shows a strong resonance at initial kinetic energy $E_{\mathrm{LO}}-E_{1e}$, any deviation from the Maxwell-Boltzmann distribution could seriously alter the average depopulation rate $g_{1e}$ - a crucial parameter for laser action. If thermalization of carriers by e-e intrasubband scattering is not significantly more efficient than intersubband scattering processes, the carriers accumulate in the subband \emph{1} up to the kinetic energy of 9 meV, in contrast to the Fermi-Dirac or the Maxwell-Boltzmann distributions\cite{Harrison_edistributionQCL_APL99,Bonno_eeMCsimuTHzQCL_JAP05}. The relative simplicity of this device would make it a good candidate for studies of electron temperature by microprobe photoluminescence experiments\cite{Vitiello_BtCTHzQCL_APL06}.

\begin{figure}[t!]
\includegraphics[width=3.5in]{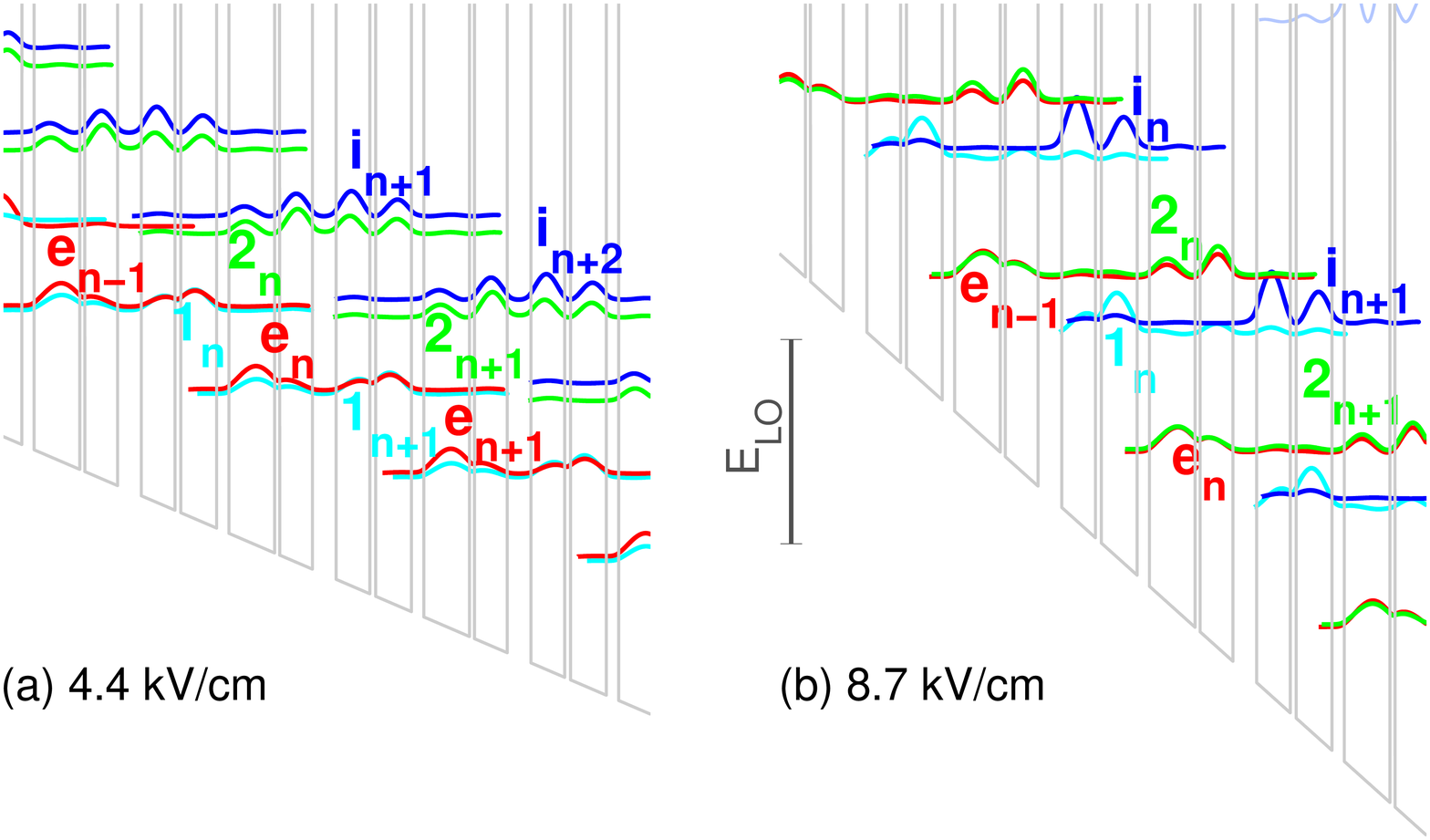}
\caption{ (Color online) Conduction band diagram and the moduli squared of wavefunctions at a) 4.4 kV/cm and b) 8.7 kV/cm.}
\label{fig_sol4p4and8p7}
\end{figure}

At 10 K, the voltage versus current density plot shows a tunneling resonance with a shoulder at $\sim$5.6 V and a strong NDR at 9.4 V (see left axis on Figure~\ref{fig_LIV}). Taking into account the 0.8 V Schottky voltage drop on the top contact, the electric field of these features correspond well to the alignment of the extraction state \emph{e} with levels \emph{1} and \emph{2}, predicted at 4.4 kV/cm and 8.7 kV/cm, respectively. Figure~\ref{fig_sol4p4and8p7} depicts the wavefunction configurations for these electric fields. The anticrossing of the lasing states occurs at 13 kV/cm, implying that below 13 kV/cm level \emph{2}(\emph{1}) is located on the downstream(upstream) side of the module. At 4.4 kV/cm, levels $\emph{e}_{\mathrm{n-1}}$ and $\emph{1}_{\mathrm{n}}$ are aligned with a coupling strength of $\Omega_{e1}=0.385~\mathrm{meV}$, while the next set of anticrossed states, $\emph{e}_{\mathrm{n}}$ and $\emph{1}_{\mathrm{n+1}}$, are 16 meV below ($qL\mathcal{E}$, $\mathcal{E}$ being the electric field). It would suggest that the transition $\emph{1}_{\mathrm{n}}\rightarrow\emph{e}_{\mathrm{n}}$ is mainly assisted by IR scattering, with a few ps lifetime. At 20 K and for the $\emph{e}\rightarrow \emph{1}$ tunneling transport $4\Omega_{e1}^{2}\tau_{\|e1}\tau_1\approx 1$, suggesting that transport is ``in between'' the coherent and incoherent regimes. Interestingly, at the same electric field, levels $\emph{2}_{\mathrm{n+1}}$ and $\emph{i}_{\mathrm{n+2}}$ anticross and are also aligned with $\emph{1}_{\mathrm{n}}$ and $\emph{e}_{\mathrm{n-1}}$. These four states could form a weakly coupled miniband (see Figure~\ref{fig_sol4p4and8p7}-a). One could imagine transport occurring two periods at a time from $\emph{e}_{\mathrm{n-1}}$ to $\emph{2}_{\mathrm{n+1}}$, since $2qL\mathcal{E}=32~\mathrm{meV}\thickapprox E_{\mathrm{LO}}$. Therefore, one can propose that the shoulder in V-J plot at $\sim$5.6 V, which is also slightly higher than \emph{e}-\emph{1} alignment voltage (4.4+0.8=5.2 V), corresponds to an electric field for which $2qL\mathcal{E}\approx\mathrm{E_{\mathrm{LO}}}$ (4.8 kV/cm when taking into account the Schottky voltage drop). At 8.7 kV/cm, levels $\emph{e}_{\mathrm{n-1}}$ and $\emph{2}_{\mathrm{n}}$ are aligned with a small coupling strength $\Omega_{e2}=0.24~\mathrm{meV}$. Transition to the next lower extraction state $\emph{e}_{\mathrm{n}}$ should be efficient, as $qL\mathcal{E}=32~\mathrm{meV}\thickapprox E_{\mathrm{LO}}$. The figure of merit of $\emph{e}\rightarrow \emph{2}$ tunneling transport is $4\Omega_{e2}^2\tau_{\|e2}\tau_2\approx 0.04\ll 1$ at 20 K, suggesting that this transport channel is incoherent, and hence very dependent on phase coherence time constant\cite{Kumar_DMmodel_PRB09,Kumar_NatPhy11}.

\begin{figure}[t!]
\centering
\includegraphics[width=3in]{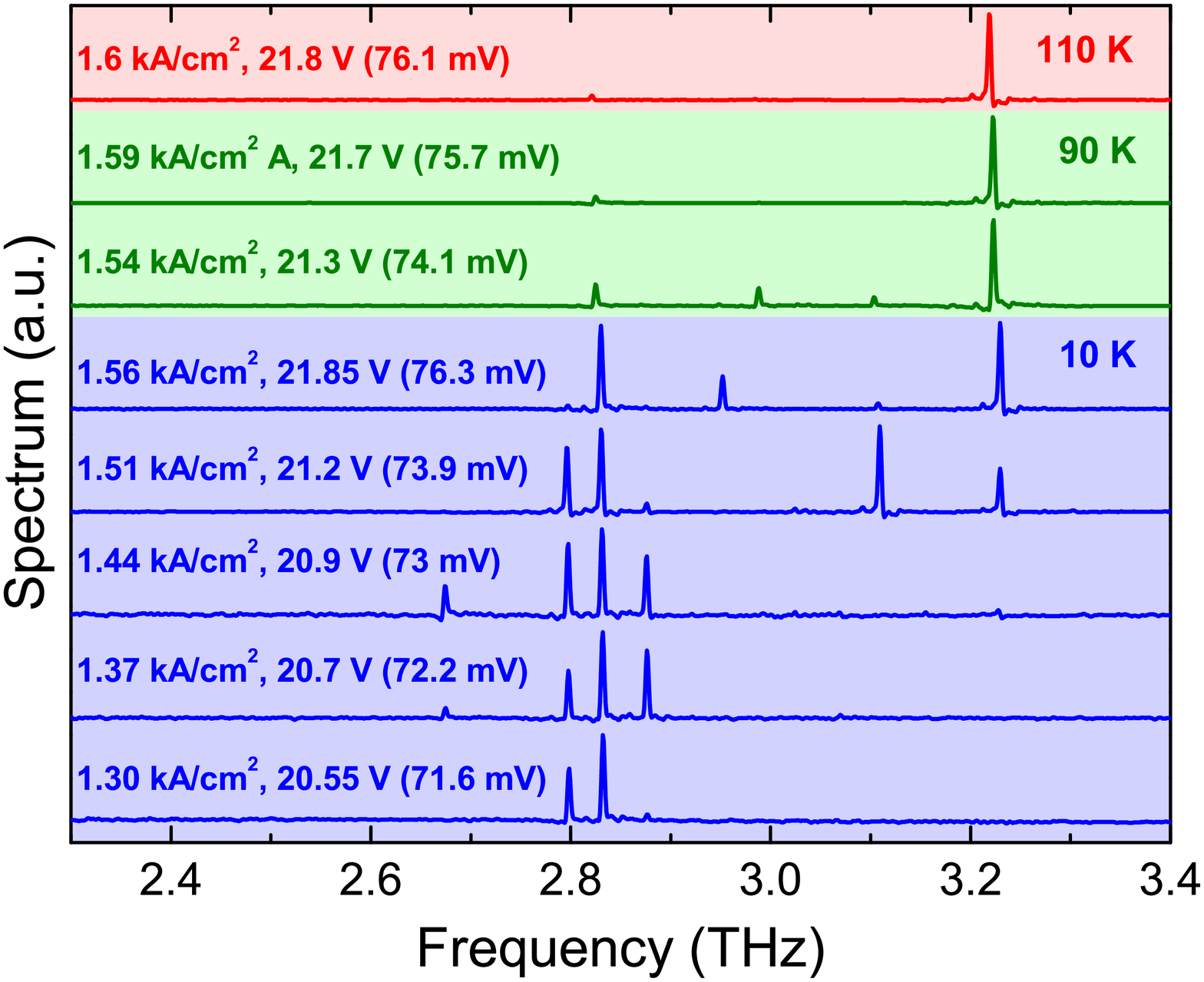}
\caption{ (Color online) THz spectra recorded for different biases and temperatures. The bias per module is reported between brackets.}
\label{fig_spectra}
\end{figure}

Figure~\ref{fig_spectra} depicts the measured spectra for the lasing device at different biases and temperatures. At 10 K and slightly above threshold, the emission line is at $\sim$2.83 THz; at higher bias, it eventually shifts to higher frequency, 3.23 THz. According to the solution of Schr\"odinger equation, in the range of electric fields for which the spectra were recorded ($\sim$19.7-21 kV/cm), the frequency spacing between the lasing states should move from 3.05 to 3.46 THz (see Figure~\ref{fig_SimuVsField}-c). At this stage, it is difficult to strongly affirm that this difference of $\sim$ 0.9 meV in photon energy is an unequivocal sign of a physical phenomenon we did take into account. However, since we have identified a signature of a significantly long LLS lifetime (no detectable discontinuity of $R_{\mathrm{ac}}$ at threshold) and therefore, a residual population on this state, we cannot exclude the occurrence of Bloch gain between the lasing states\cite{Willenberg_PRB03,Terrazi_NatPhys07}. The dispersive contribution of Bloch gain would drag the peak gain frequency to a lower value than $\nu_{21}$. This design would be an interesting candidate for gain measurements by THz time-domain spectroscopy\cite{Kroll_Nat07,Burghoff_gainM_APL11} to see if any spectral asymmetry in gain can be detected. In the next section, the results of NEGF simulations for this QCL structure are presented.

\begin{figure}[t!]
\centering
 \includegraphics[width = 3.5in]{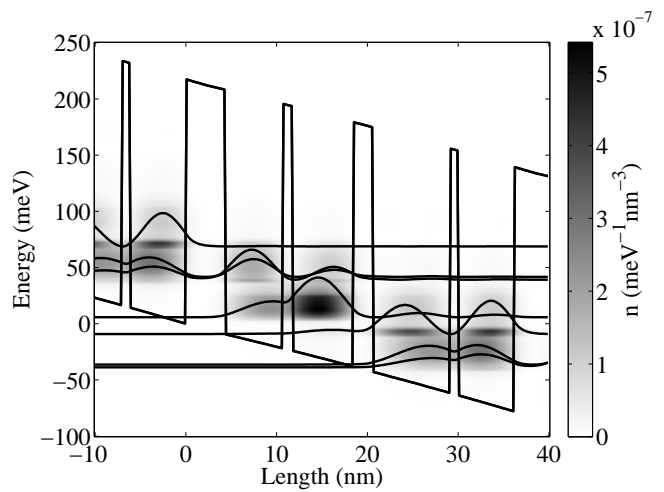}
\caption{Energy resolved electron density and the most significant basis states $|\varphi_\alpha(z)|^2$ (a. u.) at a bias of 78 mV per period and a temperature of 140 K.}
\label{fig:SimDens140K}
\end{figure}

\section{\label{NEGF} Simulations by Nonequilibrium Green's function theory}

We have simulated current density, gain spectra and carrier densities with the method of NEGF\cite{WavePacket}\cite{LeeWacker_PRB02}.
The lesser Green's function $G^<$ is related
to the density matrix $\rho_{\alpha\beta}(\mathbf{k})=-i\int \frac{dE}{2\pi}G^<_{\alpha\beta}(\mathbf{k},E)$.
The calculation is done with the basis states $\mathcal{A}^{-1/2}\sum_{\alpha} \varphi_\alpha(z)e^{-i\mathbf{k}\mathbf{r}}$, where the bold typeface
signifies two dimensional in-plane vectors and $\mathcal{A}$ is the cross-section.
This gives the energy resolved electron density
\begin{eqnarray}
n(z,E) = \frac{2}{A}\sum_\mathbf{k}\sum_{\alpha\beta} \Im\{G^<_{\alpha\beta}(\mathbf{k},E)\varphi^\ast_\beta(z)\varphi_\alpha(z) \}.
\end{eqnarray}
The current density (in A/cm${}^2$) is
\begin{eqnarray}
j(t) = \frac{1}{d}\int_0^d dz j(z,t) = \frac{2}{A}\sum_{\alpha\beta}\sum_\mathbf{k} W^{per}_{\alpha\beta}\rho_{\alpha\beta}(\mathbf{k},t),
\end{eqnarray}
where
\begin{eqnarray}
W^{per}_{\alpha\beta} \equiv \frac{1}{d}\int_0^d &&dz \frac{\hbar e}{2im(z)}\nonumber\\
&&\left(\varphi^\ast_\beta(z) \frac{\partial \varphi_\alpha(z)}{\partial z}-\frac{\partial \varphi^\ast_\beta(z)}{\partial z}\varphi_\alpha(z) \right).
\end{eqnarray}

$G^<$ is calculated on the basis of the envelope function Hamiltonian of the perfect QCL structure, where perturbations by interface roughness, ionized impurities, and phonons are treated by self-energies in the self-consistent Born approximation. The electron-electron interaction is taken into account within the mean-field approximation. In the simulations we have used an effective roughness height of 2 $\mathrm{\AA}$, and an average roughness distance of 100 $\mathrm{\AA}$. The simulated electron densities are shown in Figure~\ref{fig:SimDens140K}, along with the significant basis states $\varphi_\alpha(z)$. The darker regions close to the bottom of the subbands indicate that carriers are not thermalized, contrary to the assumption used in the previous simulations based on rate equations. We can clearly identify in which states and up to which energy carriers accumulate. This graph suggests that electrons get trapped in the LLS since the energy gap between this level and the injector state is 9 meV lower than GaAs LO-phonon energy. Moreover, we observe a slight accumulation of carriers at the bottom of the upper subband since the energy gap between this state and the ULS of the next period does not match exactly the LO-phonon energy. One can see that the shaded region associated with the extraction state is darker than that of the injection state: in other words, the population of the extraction state is significantly higher than the injection state due to the thick injection barrier. We recall the tunneling coupling strength is rather small: 1.14 meV. This makes the emptying of the extraction level slow, hence promoting the backfilling to LLS.

\begin{figure}[t!]
\centering
 \includegraphics[width = 3in]{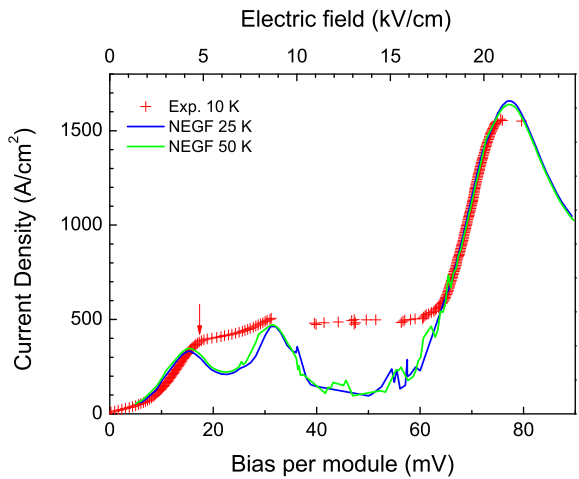}
\caption{(Color online) Current density versus voltage bias per period: experimental data (scattered points) at 10 K in pulsed mode and comparison with simulations by nonequilibrium Green's function formalism. The position of the first resonance at $\sim$4.8 kV/cm in the experimental data is highlighted by a vertical arrow and is attributed to sequential tunneling from \emph{e} to \emph{1} and then \emph{2} in three consecutive periods. The NEGF simulation are performed for higher temperatures than the experimental data but the simulated results are weakly temperature sensitive. Moreover, by the end of the current pulses (250 ns) it is well known that the lattice temperature can be several tens of kelvin higher than heat sink. The large fluctuations around 40-60 meV are artifacts of the simulation. The device was driven in current mode which explains why the valley of the second NDR predicted by theory is not observed.}
\label{fig:ivExpVsNegf}
\end{figure}

In Figure~\ref{fig:ivExpVsNegf}, we find excellent agreement with the measured data for the current densities. The second NDR at 30 mV predicted by NEGF simulations is confirmed by the experiment. The device was driven in current mode hence, on the oscilloscope, the voltage suddenly changed from 30 mV to 60 mV. The first tunneling resonance is predicted at around 4.3 kV/cm whereas the experimental current density shows a shoulder at 4.8 kV/cm (see the vertical arrow in Figure ~\ref{fig:ivExpVsNegf}) which was attributed to sequential tunneling across three modules from $\emph{e}_{\mathrm{n-1}}\rightarrow\emph{1}_{\mathrm{n}}\rightarrow\emph{2}_{\mathrm{n+1}}$ in section~\ref{ExpResults}. NEGF simulations performed around the first resonance confirmed a non negligible transport channel by phonon emission (and more marginally by other scattering sources) from the anticrossed levels \emph{2} and \emph{i}. Experimentally, the current is more phonon driven at the first resonance than predicted in NEGF simulations. For this latter model, the scattering potentials are very efficient when levels \emph{e} and \emph{1} are best aligned and therefore, give rise to a current peaked at 4.3 kV/cm.

\begin{figure}[t!]
\centering
 \includegraphics[width = 3in]{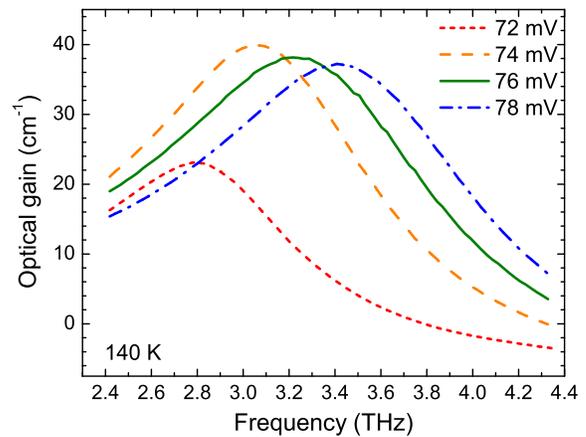}
\caption{(Color online) Gain spectra simulated using NEGF model at T=140 K and different biases per module.}
\label{fig:simuGainNEGFvsE}
\end{figure}

Finally, gain spectra have been simulated using NEGF at different temperatures and voltage biases. The results of these simulations for a fixed bias of 21.5 kV/cm (78 mV per module) and a temperature of 140 K are displayed in Figure~\ref{fig:simuGainNEGFvsE}. It is important to point out that simulations at low temperatures showed a maximum gain and peak gain frequency lower than expected from the measurements. There were numerical problems at low temperature due to the lack of e-e scattering in the simulations, whereas at 140 K the acoustic phonons assist to a higher degree in the thermalization of the carriers. An underestimated thermalization of the carriers at low temperature would enhance the trapping of electrons on the LLS, and subsequently result in a reduced population inversion and a higher contribution from Bloch gain. At 140 K, a gain of 37 $\mathrm{cm}^{-1}$ at 3.4 THz is predicted, a value that is consistent with previous measurements of waveguide loss by cavity frequency pulling\cite{Dunbar_cavityFreqPulling_APL07}. Theoretically, from 72 to 76 mV per module, the peak gain frequency moves from 2.8 to 3.24 THz, while experimentally, the laser emission shifts from 2.8 to 3.23 THz in a similar bias range (71.6 to 76.3 mV) at 10 K. This latter comparison would support the hypothesis that the simulated thermalization at 140 K by NEGF is close to the actual thermalization at lower temperatures. At 76 mV per module, the gain bandwidth derived by NEGF simulations is $\sim$1.38 THz, a value slightly broader than the assumed value of 1 THz, used in the simulations based on rate equations.

\section{\label{Conclusion} Conclusion}
This paper presented a novel THz SA-QCL design with consecutive ``phonon-photon-phonon'' emissions. The simplicity of the lasing scheme, made it convenient to calculate the energy level populations and lifetimes using a rate equation model. The first iteration structure was designed by optimizing a figure of merit defined as the product of population inversion and oscillator strength. The optimum design at 150 K only took LO-phonon scattering into account in transport calculations and assumed a Maxwell-Boltzmann distribution of carriers with a temperature of 200 K in the subbands. In the designed structure, the lasing states were fairly isolated from higher bands and not perturbed by tunneling. The fabricated device using Au double metal waveguide lased up to 138 K. The voltage -- current density characteristic of the device showed no obvious discontinuity in the differential resistance at threshold. The rate equation model suggests that this effect stems from rather slow depopulation of the LLS due to smaller-than-phonon-energy $E_{1e}$. This hypothesis was confirmed by NEGF calculations. Both rate equation and NEGF calculations indicated that the tunneling barrier is the bottleneck of the carrier transport and the carriers are piled up behind it.

Using the models discussed in this paper, we identified the shortcomings of the first iteration design and addressed them in detail for the next generations of THz SA-QCL designs. We found out that it is important to maximize the phonon scattering assisted extraction and injection rates, by a better optimization of energy spacing between corresponding subbands. Moreover, one has to take interface roughness scattering into account in the optimization process. It was found that interface roughness scattering, particularly between the lasing states, is comparable to the LO-phonon scattering rates and should not be neglected. Its effect on population dynamics and linewidth of the lasing transition also has to be considered for the next generation of THz SA-QCL designs.

The proposed lasing scheme in this work benefits from a rather simple operation principle. Consequently, such a 4-level quantum mechanical structure provides a unique platform to study experimentally various debated effects in THz QCLs, including effect of interface roughness on population dynamics and gain linewidth. The authors suggest conducting a comprehensive doping study to elucidate the debated topic of free carrier absorption and the role of e-e and e-impurity scatterings in THz QCL. Furthermore, one can explore the effect of injection tunneling barrier thickness on the coherence of the transport and its effect on the device performance. Further understanding of the carrier dynamics and gain is possible by microprobe photoluminescence experiments to study the electron distributions in the different subbands and by THz time-domain spectroscopy to investigate the possibility of Bloch gain in such a structure, respectively.

\begin{acknowledgements}
The authors would like to thank Dr. Marek Korkusinski for providing the genetic algorithm. We also would like to acknowledge the supports from Natural Science and Engineering Research Council (NSERC) of Canada, from Canadian Foundation of Innovation (CFI), from the Ontario Research Fund (ORF) and CMC Microsystems. HCL was supported in part by the National Major Basic Research Project (2011CB925603) and the Shanghai Municipal Major Basic Research Project (09DJ1400102).
\end{acknowledgements}

\nocite{*}
\bibliography{DoublePhonon_v1}

\begin{thebibliography}{43}%
\makeatletter
\providecommand \@ifxundefined [1]{%
 \@ifx{#1\undefined}
}%
\providecommand \@ifnum [1]{%
 \ifnum #1\expandafter \@firstoftwo
 \else \expandafter \@secondoftwo
 \fi
}%
\providecommand \@ifx [1]{%
 \ifx #1\expandafter \@firstoftwo
 \else \expandafter \@secondoftwo
 \fi
}%
\providecommand \natexlab [1]{#1}%
\providecommand \enquote  [1]{``#1''}%
\providecommand \bibnamefont  [1]{#1}%
\providecommand \bibfnamefont [1]{#1}%
\providecommand \citenamefont [1]{#1}%
\providecommand \href@noop [0]{\@secondoftwo}%
\providecommand \href [0]{\begingroup \@sanitize@url \@href}%
\providecommand \@href[1]{\@@startlink{#1}\@@href}%
\providecommand \@@href[1]{\endgroup#1\@@endlink}%
\providecommand \@sanitize@url [0]{\catcode `\\12\catcode `\$12\catcode
  `\&12\catcode `\#12\catcode `\^12\catcode `\_12\catcode `\%12\relax}%
\providecommand \@@startlink[1]{}%
\providecommand \@@endlink[0]{}%
\providecommand \url  [0]{\begingroup\@sanitize@url \@url }%
\providecommand \@url [1]{\endgroup\@href {#1}{\urlprefix }}%
\providecommand \urlprefix  [0]{URL }%
\providecommand \Eprint [0]{\href }%
\providecommand \doibase [0]{http://dx.doi.org/}%
\providecommand \selectlanguage [0]{\@gobble}%
\providecommand \bibinfo  [0]{\@secondoftwo}%
\providecommand \bibfield  [0]{\@secondoftwo}%
\providecommand \translation [1]{[#1]}%
\providecommand \BibitemOpen [0]{}%
\providecommand \bibitemStop [0]{}%
\providecommand \bibitemNoStop [0]{.\EOS\space}%
\providecommand \EOS [0]{\spacefactor3000\relax}%
\providecommand \BibitemShut  [1]{\csname bibitem#1\endcsname}%
\let\auto@bib@innerbib\@empty
\bibitem [{\citenamefont {Kohler}\ \emph {et~al.}(2002)\citenamefont {Kohler},
  \citenamefont {Tredicucci}, \citenamefont {Beltram}, \citenamefont {Beere},
  \citenamefont {Linfield}, \citenamefont {Davies}, \citenamefont {Ritchie},
  \citenamefont {Iotti},\ and\ \citenamefont {Rossi}}]{Kohler_Nat02}%
  \BibitemOpen
  \bibfield  {author} {\bibinfo {author} {\bibfnamefont {R.}~\bibnamefont
  {Kohler}}, \bibinfo {author} {\bibfnamefont {A.}~\bibnamefont {Tredicucci}},
  \bibinfo {author} {\bibfnamefont {F.}~\bibnamefont {Beltram}}, \bibinfo
  {author} {\bibfnamefont {H.~E.}\ \bibnamefont {Beere}}, \bibinfo {author}
  {\bibfnamefont {E.~H.}\ \bibnamefont {Linfield}}, \bibinfo {author}
  {\bibfnamefont {A.~G.}\ \bibnamefont {Davies}}, \bibinfo {author}
  {\bibfnamefont {D.}~\bibnamefont {Ritchie}}, \bibinfo {author} {\bibfnamefont
  {R.~C.}\ \bibnamefont {Iotti}}, \ and\ \bibinfo {author} {\bibfnamefont
  {F.}~\bibnamefont {Rossi}},\ }\href@noop {} {\bibfield  {journal} {\bibinfo
  {journal} {Nature}\ }\textbf {\bibinfo {volume} {417}},\ \bibinfo {pages}
  {156} (\bibinfo {year} {2002})}\BibitemShut {NoStop}%
\bibitem [{\citenamefont {Fathololoumi}\ \emph
  {et~al.}(2011{\natexlab{a}})\citenamefont {Fathololoumi}, \citenamefont
  {Dupont}, \citenamefont {Chan}, \citenamefont {Wasilewski}, \citenamefont
  {Laframboise}, \citenamefont {Ban}, \citenamefont {M\'aty\'as}, \citenamefont
  {Jirauschek}, \citenamefont {Hu},\ and\ \citenamefont
  {Liu}}]{Fathololoumi_unpub}%
  \BibitemOpen
  \bibfield  {author} {\bibinfo {author} {\bibfnamefont {S.}~\bibnamefont
  {Fathololoumi}}, \bibinfo {author} {\bibfnamefont {E.}~\bibnamefont
  {Dupont}}, \bibinfo {author} {\bibfnamefont {C.}~\bibnamefont {Chan}},
  \bibinfo {author} {\bibfnamefont {Z.}~\bibnamefont {Wasilewski}}, \bibinfo
  {author} {\bibfnamefont {S.}~\bibnamefont {Laframboise}}, \bibinfo {author}
  {\bibfnamefont {D.}~\bibnamefont {Ban}}, \bibinfo {author} {\bibfnamefont
  {A.}~\bibnamefont {M\'aty\'as}}, \bibinfo {author} {\bibfnamefont
  {C.}~\bibnamefont {Jirauschek}}, \bibinfo {author} {\bibfnamefont
  {Q.}~\bibnamefont {Hu}}, \ and\ \bibinfo {author} {\bibfnamefont {H.~C.}\
  \bibnamefont {Liu}},\ }\href@noop {} {} (\bibinfo {year}
  {2011}{\natexlab{a}}),\ \bibinfo {note} {unpublished}\BibitemShut {NoStop}%
\bibitem [{\citenamefont {Belkin}\ \emph {et~al.}(2008)\citenamefont {Belkin},
  \citenamefont {Fan}, \citenamefont {Hormoz}, \citenamefont {Capasso},
  \citenamefont {Khanna}, \citenamefont {Lachab}, \citenamefont {Davies}, ,\
  and\ \citenamefont {Linfield}}]{Belkin_OptExp_08}%
  \BibitemOpen
  \bibfield  {author} {\bibinfo {author} {\bibfnamefont {M.~A.}\ \bibnamefont
  {Belkin}}, \bibinfo {author} {\bibfnamefont {J.~A.}\ \bibnamefont {Fan}},
  \bibinfo {author} {\bibfnamefont {S.}~\bibnamefont {Hormoz}}, \bibinfo
  {author} {\bibfnamefont {F.}~\bibnamefont {Capasso}}, \bibinfo {author}
  {\bibfnamefont {S.~P.}\ \bibnamefont {Khanna}}, \bibinfo {author}
  {\bibfnamefont {M.}~\bibnamefont {Lachab}}, \bibinfo {author} {\bibfnamefont
  {A.~G.}\ \bibnamefont {Davies}}, , \ and\ \bibinfo {author} {\bibfnamefont
  {E.~H.}\ \bibnamefont {Linfield}},\ }\href@noop {} {\bibfield  {journal}
  {\bibinfo  {journal} {Opt. Exp.}\ }\textbf {\bibinfo {volume} {16}},\
  \bibinfo {pages} {3242} (\bibinfo {year} {2008})}\BibitemShut {NoStop}%
\bibitem [{\citenamefont {Kumar}, \citenamefont {Hu},\ and\ \citenamefont
  {Reno}(2009)}]{Kumar_186K_APL09}%
  \BibitemOpen
  \bibfield  {author} {\bibinfo {author} {\bibfnamefont {S.}~\bibnamefont
  {Kumar}}, \bibinfo {author} {\bibfnamefont {Q.}~\bibnamefont {Hu}}, \ and\
  \bibinfo {author} {\bibfnamefont {J.~L.}\ \bibnamefont {Reno}},\ }\href@noop
  {} {\bibfield  {journal} {\bibinfo  {journal} {Appl. Phys. Lett.}\ }\textbf
  {\bibinfo {volume} {94}},\ \bibinfo {pages} {131105} (\bibinfo {year}
  {2009})}\BibitemShut {NoStop}%
\bibitem [{\citenamefont {Belkin}\ \emph {et~al.}(2009)\citenamefont {Belkin},
  \citenamefont {Wang}, \citenamefont {Pfl\"ugl}, \citenamefont {Belyanin},
  \citenamefont {Khanna}, \citenamefont {Davies}, \citenamefont {Linfield},\
  and\ \citenamefont {Capasso}}]{Belkin_SelTopQE09}%
  \BibitemOpen
  \bibfield  {author} {\bibinfo {author} {\bibfnamefont {M.~A.}\ \bibnamefont
  {Belkin}}, \bibinfo {author} {\bibfnamefont {Q.~J.}\ \bibnamefont {Wang}},
  \bibinfo {author} {\bibfnamefont {C.}~\bibnamefont {Pfl\"ugl}}, \bibinfo
  {author} {\bibfnamefont {A.}~\bibnamefont {Belyanin}}, \bibinfo {author}
  {\bibfnamefont {S.~P.}\ \bibnamefont {Khanna}}, \bibinfo {author}
  {\bibfnamefont {A.~G.}\ \bibnamefont {Davies}}, \bibinfo {author}
  {\bibfnamefont {E.~H.}\ \bibnamefont {Linfield}}, \ and\ \bibinfo {author}
  {\bibfnamefont {F.}~\bibnamefont {Capasso}},\ }\href@noop {} {\bibfield
  {journal} {\bibinfo  {journal} {IEEE Sel. Top. Quant. Electron.}\ }\textbf
  {\bibinfo {volume} {15}},\ \bibinfo {pages} {952} (\bibinfo {year}
  {2009})}\BibitemShut {NoStop}%
\bibitem [{\citenamefont {Tonouchi}(2007)}]{Tonouchi_Nat07}%
  \BibitemOpen
  \bibfield  {author} {\bibinfo {author} {\bibfnamefont {M.}~\bibnamefont
  {Tonouchi}},\ }\href@noop {} {\bibfield  {journal} {\bibinfo  {journal} {Nat.
  Phot.}\ }\textbf {\bibinfo {volume} {1}},\ \bibinfo {pages} {97} (\bibinfo
  {year} {2007})}\BibitemShut {NoStop}%
\bibitem [{\citenamefont {Scalari}\ \emph
  {et~al.}(2007{\natexlab{a}})\citenamefont {Scalari}, \citenamefont {Terazzi},
  \citenamefont {Giovannini}, \citenamefont {Hoyler},\ and\ \citenamefont
  {Faist}}]{Scalari_APL07}%
  \BibitemOpen
  \bibfield  {author} {\bibinfo {author} {\bibfnamefont {G.}~\bibnamefont
  {Scalari}}, \bibinfo {author} {\bibfnamefont {R.}~\bibnamefont {Terazzi}},
  \bibinfo {author} {\bibfnamefont {M.}~\bibnamefont {Giovannini}}, \bibinfo
  {author} {\bibfnamefont {N.}~\bibnamefont {Hoyler}}, \ and\ \bibinfo {author}
  {\bibfnamefont {J.}~\bibnamefont {Faist}},\ }\href@noop {} {\bibfield
  {journal} {\bibinfo  {journal} {Appl. Phys. Lett.}\ }\textbf {\bibinfo
  {volume} {91}},\ \bibinfo {pages} {032103} (\bibinfo {year}
  {2007}{\natexlab{a}})}\BibitemShut {NoStop}%
\bibitem [{\citenamefont {Kumar}\ and\ \citenamefont
  {Hu}(2009)}]{Kumar_DMmodel_PRB09}%
  \BibitemOpen
  \bibfield  {author} {\bibinfo {author} {\bibfnamefont {S.}~\bibnamefont
  {Kumar}}\ and\ \bibinfo {author} {\bibfnamefont {Q.}~\bibnamefont {Hu}},\
  }\href@noop {} {\bibfield  {journal} {\bibinfo  {journal} {Phys. Rev. B}\
  }\textbf {\bibinfo {volume} {80}},\ \bibinfo {pages} {245316} (\bibinfo
  {year} {2009})}\BibitemShut {NoStop}%
\bibitem [{\citenamefont {Dupont}, \citenamefont {Fathololoumi},\ and\
  \citenamefont {Liu}(2010)}]{Dupont_PRB10}%
  \BibitemOpen
  \bibfield  {author} {\bibinfo {author} {\bibfnamefont {E.}~\bibnamefont
  {Dupont}}, \bibinfo {author} {\bibfnamefont {S.}~\bibnamefont
  {Fathololoumi}}, \ and\ \bibinfo {author} {\bibfnamefont {H.~C.}\
  \bibnamefont {Liu}},\ }\href@noop {} {\bibfield  {journal} {\bibinfo
  {journal} {Phys. Rev. B}\ }\textbf {\bibinfo {volume} {81}},\ \bibinfo
  {pages} {205311} (\bibinfo {year} {2010})}\BibitemShut {NoStop}%
\bibitem [{\citenamefont {Lee}\ and\ \citenamefont
  {Wacker}(2002)}]{LeeWacker_PRB02}%
  \BibitemOpen
  \bibfield  {author} {\bibinfo {author} {\bibfnamefont {S.~C.}\ \bibnamefont
  {Lee}}\ and\ \bibinfo {author} {\bibfnamefont {A.}~\bibnamefont {Wacker}},\
  }\href@noop {} {\bibfield  {journal} {\bibinfo  {journal} {Phys. Rev. B}\
  }\textbf {\bibinfo {volume} {66}},\ \bibinfo {pages} {245314} (\bibinfo
  {year} {2002})}\BibitemShut {NoStop}%
\bibitem [{\citenamefont {Kubis}\ \emph {et~al.}(2009)\citenamefont {Kubis},
  \citenamefont {Yeh}, \citenamefont {Vogl}, \citenamefont {Benz},
  \citenamefont {Fasching},\ and\ \citenamefont
  {Deutsch}}]{Kubis_PRB09_NEGFthzqcl}%
  \BibitemOpen
  \bibfield  {author} {\bibinfo {author} {\bibfnamefont {T.}~\bibnamefont
  {Kubis}}, \bibinfo {author} {\bibfnamefont {C.}~\bibnamefont {Yeh}}, \bibinfo
  {author} {\bibfnamefont {P.}~\bibnamefont {Vogl}}, \bibinfo {author}
  {\bibfnamefont {A.}~\bibnamefont {Benz}}, \bibinfo {author} {\bibfnamefont
  {G.}~\bibnamefont {Fasching}}, \ and\ \bibinfo {author} {\bibfnamefont
  {C.}~\bibnamefont {Deutsch}},\ }\href@noop {} {\bibfield  {journal} {\bibinfo
   {journal} {Phys. Rev. B}\ }\textbf {\bibinfo {volume} {79}},\ \bibinfo
  {pages} {195323} (\bibinfo {year} {2009})}\BibitemShut {NoStop}%
\bibitem [{\citenamefont {Schmielau}\ and\ \citenamefont
  {Pereira}(2009)}]{Schmielau_NEGF_APL09}%
  \BibitemOpen
  \bibfield  {author} {\bibinfo {author} {\bibfnamefont {T.}~\bibnamefont
  {Schmielau}}\ and\ \bibinfo {author} {\bibfnamefont {M.}~\bibnamefont
  {Pereira}},\ }\href@noop {} {\bibfield  {journal} {\bibinfo  {journal} {Appl.
  Phys. Lett}\ }\textbf {\bibinfo {volume} {95}},\ \bibinfo {pages} {231111}
  (\bibinfo {year} {2009})}\BibitemShut {NoStop}%
\bibitem [{\citenamefont {Callebaut}\ \emph {et~al.}(2003)\citenamefont
  {Callebaut}, \citenamefont {Kumar}, \citenamefont {Williams}, \citenamefont
  {Hu},\ and\ \citenamefont {Reno}}]{Callebaut_APL03}%
  \BibitemOpen
  \bibfield  {author} {\bibinfo {author} {\bibfnamefont {H.}~\bibnamefont
  {Callebaut}}, \bibinfo {author} {\bibfnamefont {S.}~\bibnamefont {Kumar}},
  \bibinfo {author} {\bibfnamefont {B.~S.}\ \bibnamefont {Williams}}, \bibinfo
  {author} {\bibfnamefont {Q.}~\bibnamefont {Hu}}, \ and\ \bibinfo {author}
  {\bibfnamefont {J.~L.}\ \bibnamefont {Reno}},\ }\href@noop {} {\bibfield
  {journal} {\bibinfo  {journal} {Appl. Phys. Lett}\ }\textbf {\bibinfo
  {volume} {83}},\ \bibinfo {pages} {207} (\bibinfo {year} {2003})}\BibitemShut
  {NoStop}%
\bibitem [{\citenamefont {Jirauschek}\ and\ \citenamefont
  {Lugli}(2009)}]{Jirauschek_JAP09_MCthzqcl}%
  \BibitemOpen
  \bibfield  {author} {\bibinfo {author} {\bibfnamefont {C.}~\bibnamefont
  {Jirauschek}}\ and\ \bibinfo {author} {\bibfnamefont {P.}~\bibnamefont
  {Lugli}},\ }\href@noop {} {\bibfield  {journal} {\bibinfo  {journal} {J.
  Appl. Phys.}\ }\textbf {\bibinfo {volume} {105}},\ \bibinfo {pages} {123102}
  (\bibinfo {year} {2009})}\BibitemShut {NoStop}%
\bibitem [{\citenamefont {Deutsch}\ \emph {et~al.}(2010)\citenamefont
  {Deutsch}, \citenamefont {Benz}, \citenamefont {Detz}, \citenamefont {Klang},
  \citenamefont {Nobile}, \citenamefont {Andrews}, \citenamefont {Schrenk},
  \citenamefont {Kubis}, \citenamefont {Vogl}, \citenamefont {Strasser},\ and\
  \citenamefont {Unterrainer}}]{Deutsch_APL10}%
  \BibitemOpen
  \bibfield  {author} {\bibinfo {author} {\bibfnamefont {C.}~\bibnamefont
  {Deutsch}}, \bibinfo {author} {\bibfnamefont {A.}~\bibnamefont {Benz}},
  \bibinfo {author} {\bibfnamefont {H.}~\bibnamefont {Detz}}, \bibinfo {author}
  {\bibfnamefont {P.}~\bibnamefont {Klang}}, \bibinfo {author} {\bibfnamefont
  {M.}~\bibnamefont {Nobile}}, \bibinfo {author} {\bibfnamefont {A.~M.}\
  \bibnamefont {Andrews}}, \bibinfo {author} {\bibfnamefont {W.}~\bibnamefont
  {Schrenk}}, \bibinfo {author} {\bibfnamefont {T.}~\bibnamefont {Kubis}},
  \bibinfo {author} {\bibfnamefont {P.}~\bibnamefont {Vogl}}, \bibinfo {author}
  {\bibfnamefont {G.}~\bibnamefont {Strasser}}, \ and\ \bibinfo {author}
  {\bibfnamefont {K.}~\bibnamefont {Unterrainer}},\ }\href@noop {} {\bibfield
  {journal} {\bibinfo  {journal} {Appl. Phys. Lett.}\ }\textbf {\bibinfo
  {volume} {97}},\ \bibinfo {pages} {261110} (\bibinfo {year}
  {2010})}\BibitemShut {NoStop}%
\bibitem [{\citenamefont {Yasuda}\ \emph {et~al.}(2009)\citenamefont {Yasuda},
  \citenamefont {Kubis}, \citenamefont {Vogl}, \citenamefont {Sekine},
  \citenamefont {Hosako},\ and\ \citenamefont {Hirakawa}}]{Yasuda_APL09}%
  \BibitemOpen
  \bibfield  {author} {\bibinfo {author} {\bibfnamefont {H.}~\bibnamefont
  {Yasuda}}, \bibinfo {author} {\bibfnamefont {T.}~\bibnamefont {Kubis}},
  \bibinfo {author} {\bibfnamefont {P.}~\bibnamefont {Vogl}}, \bibinfo {author}
  {\bibfnamefont {N.}~\bibnamefont {Sekine}}, \bibinfo {author} {\bibfnamefont
  {I.}~\bibnamefont {Hosako}}, \ and\ \bibinfo {author} {\bibfnamefont
  {K.}~\bibnamefont {Hirakawa}},\ }\href@noop {} {\bibfield  {journal}
  {\bibinfo  {journal} {Appl. Phys. Lett.}\ }\textbf {\bibinfo {volume} {94}},\
  \bibinfo {pages} {151109} (\bibinfo {year} {2009})}\BibitemShut {NoStop}%
\bibitem [{\citenamefont {Kubis}, \citenamefont {Mehrotra},\ and\ \citenamefont
  {Klimeck}(2010)}]{Kubis_APL10}%
  \BibitemOpen
  \bibfield  {author} {\bibinfo {author} {\bibfnamefont {T.}~\bibnamefont
  {Kubis}}, \bibinfo {author} {\bibfnamefont {S.~R.}\ \bibnamefont {Mehrotra}},
  \ and\ \bibinfo {author} {\bibfnamefont {G.}~\bibnamefont {Klimeck}},\
  }\href@noop {} {\bibfield  {journal} {\bibinfo  {journal} {Appl. Phys.
  Lett.}\ }\textbf {\bibinfo {volume} {97}},\ \bibinfo {pages} {261106}
  (\bibinfo {year} {2010})}\BibitemShut {NoStop}%
\bibitem [{\citenamefont {Kumar}\ \emph {et~al.}(2011)\citenamefont {Kumar},
  \citenamefont {Chan}, \citenamefont {Hu},\ and\ \citenamefont
  {Reno}}]{Kumar_NatPhy11}%
  \BibitemOpen
  \bibfield  {author} {\bibinfo {author} {\bibfnamefont {S.}~\bibnamefont
  {Kumar}}, \bibinfo {author} {\bibfnamefont {C.~W.~I.}\ \bibnamefont {Chan}},
  \bibinfo {author} {\bibfnamefont {Q.}~\bibnamefont {Hu}}, \ and\ \bibinfo
  {author} {\bibfnamefont {J.~L.}\ \bibnamefont {Reno}},\ }\href@noop {}
  {\bibfield  {journal} {\bibinfo  {journal} {Nat. Phys.}\ }\textbf {\bibinfo
  {volume} {7}},\ \bibinfo {pages} {166} (\bibinfo {year} {2011})}\BibitemShut
  {NoStop}%
\bibitem [{\citenamefont {Yamanishi}\ \emph {et~al.}(2008)\citenamefont
  {Yamanishi}, \citenamefont {Fujita}, \citenamefont {Edamura},\ and\
  \citenamefont {Kan}}]{Yamanishi_OptExp_08}%
  \BibitemOpen
  \bibfield  {author} {\bibinfo {author} {\bibfnamefont {M.}~\bibnamefont
  {Yamanishi}}, \bibinfo {author} {\bibfnamefont {K.}~\bibnamefont {Fujita}},
  \bibinfo {author} {\bibfnamefont {T.}~\bibnamefont {Edamura}}, \ and\
  \bibinfo {author} {\bibfnamefont {H.}~\bibnamefont {Kan}},\ }\href@noop {}
  {\bibfield  {journal} {\bibinfo  {journal} {Opt. Exp.}\ }\textbf {\bibinfo
  {volume} {16}},\ \bibinfo {pages} {20748} (\bibinfo {year}
  {2008})}\BibitemShut {NoStop}%
\bibitem [{\citenamefont {Yamanishi}\ \emph {et~al.}(2011)\citenamefont
  {Yamanishi}, \citenamefont {Fujita}, \citenamefont {Kubis}, \citenamefont
  {Yu}, \citenamefont {Edamura}, \citenamefont {Tanaka}, \citenamefont
  {Klimeck},\ and\ \citenamefont {Capasso}}]{Yamanishi_ITQW11}%
  \BibitemOpen
  \bibfield  {author} {\bibinfo {author} {\bibfnamefont {M.}~\bibnamefont
  {Yamanishi}}, \bibinfo {author} {\bibfnamefont {K.}~\bibnamefont {Fujita}},
  \bibinfo {author} {\bibfnamefont {T.}~\bibnamefont {Kubis}}, \bibinfo
  {author} {\bibfnamefont {N.}~\bibnamefont {Yu}}, \bibinfo {author}
  {\bibfnamefont {T.}~\bibnamefont {Edamura}}, \bibinfo {author} {\bibfnamefont
  {K.}~\bibnamefont {Tanaka}}, \bibinfo {author} {\bibfnamefont
  {G.}~\bibnamefont {Klimeck}}, \ and\ \bibinfo {author} {\bibfnamefont
  {F.}~\bibnamefont {Capasso}}\ }(\bibinfo {address} {presented at Eleventh
  International Conference on Intersubband Transitions in Quantum Wells,
  Badesi, Italy},\ \bibinfo {year} {2011})\BibitemShut {NoStop}%
\bibitem [{\citenamefont {Nelander}\ and\ \citenamefont
  {Wacker}(2008)}]{NelanderWacker_APL08}%
  \BibitemOpen
  \bibfield  {author} {\bibinfo {author} {\bibfnamefont {R.}~\bibnamefont
  {Nelander}}\ and\ \bibinfo {author} {\bibfnamefont {A.}~\bibnamefont
  {Wacker}},\ }\href@noop {} {\bibfield  {journal} {\bibinfo  {journal} {Appl.
  Phys. Lett.}\ }\textbf {\bibinfo {volume} {92}},\ \bibinfo {pages} {081102}
  (\bibinfo {year} {2008})}\BibitemShut {NoStop}%
\bibitem [{\citenamefont {Wacker}, \citenamefont {Nelander},\ and\
  \citenamefont {Weber}(2009)}]{Wacker_SPIE09_gainQCL}%
  \BibitemOpen
  \bibfield  {author} {\bibinfo {author} {\bibfnamefont {A.}~\bibnamefont
  {Wacker}}, \bibinfo {author} {\bibfnamefont {R.}~\bibnamefont {Nelander}}, \
  and\ \bibinfo {author} {\bibfnamefont {C.}~\bibnamefont {Weber}},\ }in\
  \href@noop {} {\emph {\bibinfo {booktitle} {Novel In-Plane Semiconductor
  Lasers VIII}}},\ Vol.\ \bibinfo {volume} {7230},\ \bibinfo {editor} {edited
  by\ \bibinfo {editor} {\bibfnamefont {A.~A.}\ \bibnamefont {Belyanin}}\ and\
  \bibinfo {editor} {\bibfnamefont {P.~M.}\ \bibnamefont {Smowton}}}\ (\bibinfo
  {organization} {SPIE},\ \bibinfo {address} {Bellingham, WA},\ \bibinfo {year}
  {2009})\ p.\ \bibinfo {pages} {72301A}\BibitemShut {NoStop}%
\bibitem [{\citenamefont {Scalari}\ \emph {et~al.}(2010)\citenamefont
  {Scalari}, \citenamefont {Amanti}, \citenamefont {Terazzi}, \citenamefont
  {Beck}, \citenamefont {Walther},\ and\ \citenamefont
  {Faist}}]{Scalari_DoubWell_OE10}%
  \BibitemOpen
  \bibfield  {author} {\bibinfo {author} {\bibfnamefont {G.}~\bibnamefont
  {Scalari}}, \bibinfo {author} {\bibfnamefont {M.~I.}\ \bibnamefont {Amanti}},
  \bibinfo {author} {\bibfnamefont {R.}~\bibnamefont {Terazzi}}, \bibinfo
  {author} {\bibfnamefont {M.}~\bibnamefont {Beck}}, \bibinfo {author}
  {\bibfnamefont {C.}~\bibnamefont {Walther}}, \ and\ \bibinfo {author}
  {\bibfnamefont {J.}~\bibnamefont {Faist}},\ }\href@noop {} {\bibfield
  {journal} {\bibinfo  {journal} {Opt. Express}\ }\textbf {\bibinfo {volume}
  {7}},\ \bibinfo {pages} {8043} (\bibinfo {year} {2010})}\BibitemShut
  {NoStop}%
\bibitem [{\citenamefont {Fathololoumi}\ \emph
  {et~al.}(2011{\natexlab{b}})\citenamefont {Fathololoumi}, \citenamefont
  {Dupont}, \citenamefont {Wasilewski}, \citenamefont {Laframboise},
  \citenamefont {Ban},\ and\ \citenamefont {Liu}}]{Fathololoumi_ITQW11}%
  \BibitemOpen
  \bibfield  {author} {\bibinfo {author} {\bibfnamefont {S.}~\bibnamefont
  {Fathololoumi}}, \bibinfo {author} {\bibfnamefont {E.}~\bibnamefont
  {Dupont}}, \bibinfo {author} {\bibfnamefont {Z.~R.}\ \bibnamefont
  {Wasilewski}}, \bibinfo {author} {\bibfnamefont {S.~R.}\ \bibnamefont
  {Laframboise}}, \bibinfo {author} {\bibfnamefont {D.}~\bibnamefont {Ban}}, \
  and\ \bibinfo {author} {\bibfnamefont {H.~C.}\ \bibnamefont {Liu}}\
  }(\bibinfo {address} {presented at Eleventh International Conference on
  Intersubband Transitions in Quantum Wells, Badesi, Italy},\ \bibinfo {year}
  {2011})\BibitemShut {NoStop}%
\bibitem [{\citenamefont {Kumar}\ \emph {et~al.}(2009)\citenamefont {Kumar},
  \citenamefont {Chan}, \citenamefont {Hu},\ and\ \citenamefont
  {Reno}}]{Kumar_2well_APL09}%
  \BibitemOpen
  \bibfield  {author} {\bibinfo {author} {\bibfnamefont {S.}~\bibnamefont
  {Kumar}}, \bibinfo {author} {\bibfnamefont {C.~W.~I.}\ \bibnamefont {Chan}},
  \bibinfo {author} {\bibfnamefont {Q.}~\bibnamefont {Hu}}, \ and\ \bibinfo
  {author} {\bibfnamefont {J.~L.}\ \bibnamefont {Reno}},\ }\href@noop {}
  {\bibfield  {journal} {\bibinfo  {journal} {Appl. Phys. Lett.}\ }\textbf
  {\bibinfo {volume} {95}},\ \bibinfo {pages} {141110} (\bibinfo {year}
  {2009})}\BibitemShut {NoStop}%
\bibitem [{\citenamefont {Wacker}(2010)}]{Wacker_APL10}%
  \BibitemOpen
  \bibfield  {author} {\bibinfo {author} {\bibfnamefont {A.}~\bibnamefont
  {Wacker}},\ }\href@noop {} {\bibfield  {journal} {\bibinfo  {journal} {Appl.
  Phys. Lett.}\ }\textbf {\bibinfo {volume} {97}},\ \bibinfo {pages} {081105}
  (\bibinfo {year} {2010})}\BibitemShut {NoStop}%
\bibitem [{\citenamefont {Khurgin}\ and\ \citenamefont
  {Dikmelik}(2010)}]{Khurgin_rateequation10}%
  \BibitemOpen
  \bibfield  {author} {\bibinfo {author} {\bibfnamefont {J.~B.}\ \bibnamefont
  {Khurgin}}\ and\ \bibinfo {author} {\bibfnamefont {Y.}~\bibnamefont
  {Dikmelik}},\ }\href@noop {} {\bibfield  {journal} {\bibinfo  {journal} {Opt.
  Engin.}\ }\textbf {\bibinfo {volume} {49}},\ \bibinfo {pages} {111110}
  (\bibinfo {year} {2010})}\BibitemShut {NoStop}%
\bibitem [{\citenamefont {Carosella}\ \emph {et~al.}(2011)\citenamefont
  {Carosella}, \citenamefont {Ndebeka-Bandou}, \citenamefont {Ferreira},
  \citenamefont {Dupont}, \citenamefont {Unterrainer}, \citenamefont
  {Strasser}, \citenamefont {Wacker},\ and\ \citenamefont
  {Bastard}}]{Carosella_FCA_arXiv}%
  \BibitemOpen
  \bibfield  {author} {\bibinfo {author} {\bibfnamefont {F.}~\bibnamefont
  {Carosella}}, \bibinfo {author} {\bibfnamefont {C.}~\bibnamefont
  {Ndebeka-Bandou}}, \bibinfo {author} {\bibfnamefont {R.}~\bibnamefont
  {Ferreira}}, \bibinfo {author} {\bibfnamefont {E.}~\bibnamefont {Dupont}},
  \bibinfo {author} {\bibfnamefont {K.}~\bibnamefont {Unterrainer}}, \bibinfo
  {author} {\bibfnamefont {S.}~\bibnamefont {Strasser}}, \bibinfo {author}
  {\bibfnamefont {A.}~\bibnamefont {Wacker}}, \ and\ \bibinfo {author}
  {\bibfnamefont {G.}~\bibnamefont {Bastard}},\ }\href@noop {} {} (\bibinfo
  {year} {2011}),\ \Eprint {http://arxiv.org/abs/1112.1822v1
  [cond-mat.mes-hall]} {arXiv:1112.1822v1 [cond-mat.mes-hall]} \BibitemShut
  {NoStop}%
\bibitem [{\citenamefont {Scalari}\ \emph
  {et~al.}(2007{\natexlab{b}})\citenamefont {Scalari}, \citenamefont {Walther},
  \citenamefont {Sirigu}, \citenamefont {Sadowski}, \citenamefont {Beere},
  \citenamefont {Ritchie}, \citenamefont {Hoyler}, \citenamefont {Giovannini},\
  and\ \citenamefont {Faist}}]{Scalari_THZQCL_Hfield_PRB07}%
  \BibitemOpen
  \bibfield  {author} {\bibinfo {author} {\bibfnamefont {G.}~\bibnamefont
  {Scalari}}, \bibinfo {author} {\bibfnamefont {C.}~\bibnamefont {Walther}},
  \bibinfo {author} {\bibfnamefont {L.}~\bibnamefont {Sirigu}}, \bibinfo
  {author} {\bibfnamefont {M.~L.}\ \bibnamefont {Sadowski}}, \bibinfo {author}
  {\bibfnamefont {H.}~\bibnamefont {Beere}}, \bibinfo {author} {\bibfnamefont
  {D.}~\bibnamefont {Ritchie}}, \bibinfo {author} {\bibfnamefont
  {N.}~\bibnamefont {Hoyler}}, \bibinfo {author} {\bibfnamefont
  {M.}~\bibnamefont {Giovannini}}, \ and\ \bibinfo {author} {\bibfnamefont
  {J.}~\bibnamefont {Faist}},\ }\href@noop {} {\bibfield  {journal} {\bibinfo
  {journal} {Phys. Rev. B}\ }\textbf {\bibinfo {volume} {76}},\ \bibinfo
  {pages} {115305} (\bibinfo {year} {2007}{\natexlab{b}})}\BibitemShut
  {NoStop}%
\bibitem [{\citenamefont {Unuma}\ \emph {et~al.}(2001)\citenamefont {Unuma},
  \citenamefont {Takahashi}, \citenamefont {Noda}, \citenamefont {Yoshita},
  \citenamefont {Sakaki}, \citenamefont {Baba},\ and\ \citenamefont
  {Akiyama}}]{Unuma_IR_APL01}%
  \BibitemOpen
  \bibfield  {author} {\bibinfo {author} {\bibfnamefont {T.}~\bibnamefont
  {Unuma}}, \bibinfo {author} {\bibfnamefont {T.}~\bibnamefont {Takahashi}},
  \bibinfo {author} {\bibfnamefont {T.}~\bibnamefont {Noda}}, \bibinfo {author}
  {\bibfnamefont {M.}~\bibnamefont {Yoshita}}, \bibinfo {author} {\bibfnamefont
  {H.}~\bibnamefont {Sakaki}}, \bibinfo {author} {\bibfnamefont
  {M.}~\bibnamefont {Baba}}, \ and\ \bibinfo {author} {\bibfnamefont
  {H.}~\bibnamefont {Akiyama}},\ }\href@noop {} {\bibfield  {journal} {\bibinfo
   {journal} {Appl. Phys. Lett.}\ }\textbf {\bibinfo {volume} {78}},\ \bibinfo
  {pages} {3448} (\bibinfo {year} {2001})}\BibitemShut {NoStop}%
\bibitem [{\citenamefont {Tsujino}\ \emph {et~al.}(2005)\citenamefont
  {Tsujino}, \citenamefont {Borak}, \citenamefont {M\"uller}, \citenamefont
  {Scheinert}, \citenamefont {Falub}, \citenamefont {Sigg}, \citenamefont
  {Gr\"utzmacher}, \citenamefont {Giovannini},\ and\ \citenamefont
  {Faist}}]{Tsujino_APL05}%
  \BibitemOpen
  \bibfield  {author} {\bibinfo {author} {\bibfnamefont {S.}~\bibnamefont
  {Tsujino}}, \bibinfo {author} {\bibfnamefont {A.}~\bibnamefont {Borak}},
  \bibinfo {author} {\bibfnamefont {E.}~\bibnamefont {M\"uller}}, \bibinfo
  {author} {\bibfnamefont {M.}~\bibnamefont {Scheinert}}, \bibinfo {author}
  {\bibfnamefont {C.~V.}\ \bibnamefont {Falub}}, \bibinfo {author}
  {\bibfnamefont {H.}~\bibnamefont {Sigg}}, \bibinfo {author} {\bibfnamefont
  {D.}~\bibnamefont {Gr\"utzmacher}}, \bibinfo {author} {\bibfnamefont
  {M.}~\bibnamefont {Giovannini}}, \ and\ \bibinfo {author} {\bibfnamefont
  {J.}~\bibnamefont {Faist}},\ }\href@noop {} {\bibfield  {journal} {\bibinfo
  {journal} {Appl. Phys. Lett.}\ }\textbf {\bibinfo {volume} {86}},\ \bibinfo
  {pages} {062113} (\bibinfo {year} {2005})}\BibitemShut {NoStop}%
\bibitem [{\citenamefont {Wasilewski}, \citenamefont {Wu},\ and\ \citenamefont
  {Dupont}()}]{Wasilewski_1stperiodQCL_unpub}%
  \BibitemOpen
  \bibfield  {author} {\bibinfo {author} {\bibfnamefont {Z.}~\bibnamefont
  {Wasilewski}}, \bibinfo {author} {\bibfnamefont {X.}~\bibnamefont {Wu}}, \
  and\ \bibinfo {author} {\bibfnamefont {E.}~\bibnamefont {Dupont}},\
  }\href@noop {} {}\bibinfo {note} {(private communication)}\BibitemShut
  {NoStop}%
\bibitem [{\citenamefont {Williams}\ \emph {et~al.}(2003)\citenamefont
  {Williams}, \citenamefont {Kumar}, \citenamefont {Callebaut}, \citenamefont
  {Hu},\ and\ \citenamefont {Reno}}]{Williams_MMwaveguide_APL03}%
  \BibitemOpen
  \bibfield  {author} {\bibinfo {author} {\bibfnamefont {B.~S.}\ \bibnamefont
  {Williams}}, \bibinfo {author} {\bibfnamefont {S.}~\bibnamefont {Kumar}},
  \bibinfo {author} {\bibfnamefont {H.}~\bibnamefont {Callebaut}}, \bibinfo
  {author} {\bibfnamefont {Q.}~\bibnamefont {Hu}}, \ and\ \bibinfo {author}
  {\bibfnamefont {J.~L.}\ \bibnamefont {Reno}},\ }\href@noop {} {\bibfield
  {journal} {\bibinfo  {journal} {Appl. Phys. Lett.}\ }\textbf {\bibinfo
  {volume} {83}},\ \bibinfo {pages} {2124} (\bibinfo {year}
  {2003})}\BibitemShut {NoStop}%
\bibitem [{\citenamefont {Fathololoumi}\ \emph
  {et~al.}(2011{\natexlab{c}})\citenamefont {Fathololoumi}, \citenamefont
  {Dupont}, \citenamefont {Razavipour}, \citenamefont {Laframboise},
  \citenamefont {Parent}, \citenamefont {Wasilewski}, \citenamefont {Liu},\
  and\ \citenamefont {Ban}}]{Fathololoumi_SST11}%
  \BibitemOpen
  \bibfield  {author} {\bibinfo {author} {\bibfnamefont {S.}~\bibnamefont
  {Fathololoumi}}, \bibinfo {author} {\bibfnamefont {E.}~\bibnamefont
  {Dupont}}, \bibinfo {author} {\bibfnamefont {S.}~\bibnamefont {Razavipour}},
  \bibinfo {author} {\bibfnamefont {S.~R.}\ \bibnamefont {Laframboise}},
  \bibinfo {author} {\bibfnamefont {G.}~\bibnamefont {Parent}}, \bibinfo
  {author} {\bibfnamefont {Z.}~\bibnamefont {Wasilewski}}, \bibinfo {author}
  {\bibfnamefont {H.~C.}\ \bibnamefont {Liu}}, \ and\ \bibinfo {author}
  {\bibfnamefont {D.}~\bibnamefont {Ban}},\ }\href@noop {} {\bibfield
  {journal} {\bibinfo  {journal} {Semicond. Sci. Technol.}\ }\textbf {\bibinfo
  {volume} {26}},\ \bibinfo {pages} {105021} (\bibinfo {year}
  {2011}{\natexlab{c}})}\BibitemShut {NoStop}%
\bibitem [{\citenamefont {Harrison}(1999)}]{Harrison_edistributionQCL_APL99}%
  \BibitemOpen
  \bibfield  {author} {\bibinfo {author} {\bibfnamefont {P.}~\bibnamefont
  {Harrison}},\ }\href@noop {} {\bibfield  {journal} {\bibinfo  {journal}
  {Appl. Phys. Lett.}\ }\textbf {\bibinfo {volume} {75}},\ \bibinfo {pages}
  {2800} (\bibinfo {year} {1999})}\BibitemShut {NoStop}%
\bibitem [{\citenamefont {Bonno}, \citenamefont {Thobel},\ and\ \citenamefont
  {Dessenne}(2005)}]{Bonno_eeMCsimuTHzQCL_JAP05}%
  \BibitemOpen
  \bibfield  {author} {\bibinfo {author} {\bibfnamefont {O.}~\bibnamefont
  {Bonno}}, \bibinfo {author} {\bibfnamefont {J.-L.}\ \bibnamefont {Thobel}}, \
  and\ \bibinfo {author} {\bibfnamefont {F.}~\bibnamefont {Dessenne}},\
  }\href@noop {} {\bibfield  {journal} {\bibinfo  {journal} {J. Appl. Phys.}\
  }\textbf {\bibinfo {volume} {97}},\ \bibinfo {pages} {043702} (\bibinfo
  {year} {2005})}\BibitemShut {NoStop}%
\bibitem [{\citenamefont {Vitiello}\ \emph {et~al.}(2006)\citenamefont
  {Vitiello}, \citenamefont {Scarmacio}, \citenamefont {Spagnolo},
  \citenamefont {Losco}, \citenamefont {green}, \citenamefont {Tredicucci},
  \citenamefont {Beere},\ and\ \citenamefont
  {Ritchie}}]{Vitiello_BtCTHzQCL_APL06}%
  \BibitemOpen
  \bibfield  {author} {\bibinfo {author} {\bibfnamefont {M.~S.}\ \bibnamefont
  {Vitiello}}, \bibinfo {author} {\bibfnamefont {G.}~\bibnamefont {Scarmacio}},
  \bibinfo {author} {\bibfnamefont {V.}~\bibnamefont {Spagnolo}}, \bibinfo
  {author} {\bibfnamefont {T.}~\bibnamefont {Losco}}, \bibinfo {author}
  {\bibfnamefont {R.~P.}\ \bibnamefont {green}}, \bibinfo {author}
  {\bibfnamefont {A.}~\bibnamefont {Tredicucci}}, \bibinfo {author}
  {\bibfnamefont {H.~E.}\ \bibnamefont {Beere}}, \ and\ \bibinfo {author}
  {\bibfnamefont {D.~A.}\ \bibnamefont {Ritchie}},\ }\href@noop {} {\bibfield
  {journal} {\bibinfo  {journal} {Appl. Phys. Lett.}\ }\textbf {\bibinfo
  {volume} {88}},\ \bibinfo {pages} {241109} (\bibinfo {year}
  {2006})}\BibitemShut {NoStop}%
\bibitem [{\citenamefont {Willenberg}, \citenamefont {D\"ohler},\ and\
  \citenamefont {Faist}(2003)}]{Willenberg_PRB03}%
  \BibitemOpen
  \bibfield  {author} {\bibinfo {author} {\bibfnamefont {H.}~\bibnamefont
  {Willenberg}}, \bibinfo {author} {\bibfnamefont {G.~H.}\ \bibnamefont
  {D\"ohler}}, \ and\ \bibinfo {author} {\bibfnamefont {J.}~\bibnamefont
  {Faist}},\ }\href@noop {} {\bibfield  {journal} {\bibinfo  {journal} {Phys.
  Rev. B}\ }\textbf {\bibinfo {volume} {67}},\ \bibinfo {pages} {085315}
  (\bibinfo {year} {2003})}\BibitemShut {NoStop}%
\bibitem [{\citenamefont {Terazzi}\ \emph {et~al.}(2007)\citenamefont
  {Terazzi}, \citenamefont {Gresch}, \citenamefont {Giovannini}, \citenamefont
  {Hoyler}, \citenamefont {Sekine},\ and\ \citenamefont
  {Faist}}]{Terrazi_NatPhys07}%
  \BibitemOpen
  \bibfield  {author} {\bibinfo {author} {\bibfnamefont {R.}~\bibnamefont
  {Terazzi}}, \bibinfo {author} {\bibfnamefont {T.}~\bibnamefont {Gresch}},
  \bibinfo {author} {\bibfnamefont {M.}~\bibnamefont {Giovannini}}, \bibinfo
  {author} {\bibfnamefont {N.}~\bibnamefont {Hoyler}}, \bibinfo {author}
  {\bibfnamefont {N.}~\bibnamefont {Sekine}}, \ and\ \bibinfo {author}
  {\bibfnamefont {J.}~\bibnamefont {Faist}},\ }\href@noop {} {\bibfield
  {journal} {\bibinfo  {journal} {Nat. Phys.}\ }\textbf {\bibinfo {volume}
  {3}},\ \bibinfo {pages} {329} (\bibinfo {year} {2007})}\BibitemShut {NoStop}%
\bibitem [{\citenamefont {Kr\"oll}\ \emph {et~al.}(2007)\citenamefont
  {Kr\"oll}, \citenamefont {Darmo}, \citenamefont {Dhillon}, \citenamefont
  {Marcadet}, \citenamefont {Calligaro}, \citenamefont {Sirtori},\ and\
  \citenamefont {Unterrainer}}]{Kroll_Nat07}%
  \BibitemOpen
  \bibfield  {author} {\bibinfo {author} {\bibfnamefont {J.}~\bibnamefont
  {Kr\"oll}}, \bibinfo {author} {\bibfnamefont {J.}~\bibnamefont {Darmo}},
  \bibinfo {author} {\bibfnamefont {S.~S.}\ \bibnamefont {Dhillon}}, \bibinfo
  {author} {\bibfnamefont {X.}~\bibnamefont {Marcadet}}, \bibinfo {author}
  {\bibfnamefont {M.}~\bibnamefont {Calligaro}}, \bibinfo {author}
  {\bibfnamefont {C.}~\bibnamefont {Sirtori}}, \ and\ \bibinfo {author}
  {\bibfnamefont {K.}~\bibnamefont {Unterrainer}},\ }\href@noop {} {\bibfield
  {journal} {\bibinfo  {journal} {Nature}\ }\textbf {\bibinfo {volume} {449}}
  (\bibinfo {year} {2007})}\BibitemShut {NoStop}%
\bibitem [{\citenamefont {Burghoff}\ \emph {et~al.}(2011)\citenamefont
  {Burghoff}, \citenamefont {Kao}, \citenamefont {Ban}, \citenamefont {Lee},
  \citenamefont {Hu},\ and\ \citenamefont {Reno}}]{Burghoff_gainM_APL11}%
  \BibitemOpen
  \bibfield  {author} {\bibinfo {author} {\bibfnamefont {D.}~\bibnamefont
  {Burghoff}}, \bibinfo {author} {\bibfnamefont {T.-Y.}\ \bibnamefont {Kao}},
  \bibinfo {author} {\bibfnamefont {D.}~\bibnamefont {Ban}}, \bibinfo {author}
  {\bibfnamefont {A.~W.~M.}\ \bibnamefont {Lee}}, \bibinfo {author}
  {\bibfnamefont {Q.}~\bibnamefont {Hu}}, \ and\ \bibinfo {author}
  {\bibfnamefont {J.}~\bibnamefont {Reno}},\ }\href@noop {} {\bibfield
  {journal} {\bibinfo  {journal} {Appl. Phys. Lett.}\ }\textbf {\bibinfo
  {volume} {98}},\ \bibinfo {pages} {061112} (\bibinfo {year}
  {2011})}\BibitemShut {NoStop}%
\bibitem [{\citenamefont {Lee}\ \emph {et~al.}(2006)\citenamefont {Lee},
  \citenamefont {Banit}, \citenamefont {Woerner},\ and\ \citenamefont
  {Wacker}}]{WavePacket}%
  \BibitemOpen
  \bibfield  {author} {\bibinfo {author} {\bibfnamefont {C.-S.}\ \bibnamefont
  {Lee}}, \bibinfo {author} {\bibfnamefont {F.}~\bibnamefont {Banit}}, \bibinfo
  {author} {\bibfnamefont {M.}~\bibnamefont {Woerner}}, \ and\ \bibinfo
  {author} {\bibfnamefont {A.}~\bibnamefont {Wacker}},\ }\href@noop {}
  {\bibfield  {journal} {\bibinfo  {journal} {Phys. Rev. B}\ }\textbf {\bibinfo
  {volume} {73}},\ \bibinfo {pages} {245320} (\bibinfo {year}
  {2006})}\BibitemShut {NoStop}%
\bibitem [{\citenamefont {Dunbar}\ \emph {et~al.}(2007)\citenamefont {Dunbar},
  \citenamefont {Houdr\'e}, \citenamefont {Scalari}, \citenamefont {Sirugu},
  \citenamefont {Giovannini},\ and\ \citenamefont
  {Faist}}]{Dunbar_cavityFreqPulling_APL07}%
  \BibitemOpen
  \bibfield  {author} {\bibinfo {author} {\bibfnamefont {L.~A.}\ \bibnamefont
  {Dunbar}}, \bibinfo {author} {\bibfnamefont {R.}~\bibnamefont {Houdr\'e}},
  \bibinfo {author} {\bibfnamefont {G.}~\bibnamefont {Scalari}}, \bibinfo
  {author} {\bibfnamefont {L.}~\bibnamefont {Sirugu}}, \bibinfo {author}
  {\bibfnamefont {M.}~\bibnamefont {Giovannini}}, \ and\ \bibinfo {author}
  {\bibfnamefont {J.}~\bibnamefont {Faist}},\ }\href@noop {} {\bibfield
  {journal} {\bibinfo  {journal} {Appl. Phys. Lett.}\ }\textbf {\bibinfo
  {volume} {90}},\ \bibinfo {pages} {141114} (\bibinfo {year}
  {2007})}\BibitemShut {NoStop}%
\end{thebibliography}%
\end{document}